\documentclass[reprint,preprintnumbers,superscriptaddress,amsmath,amssymb,prb]{revtex4-1}
\usepackage[T1]{fontenc} 
\usepackage[utf8]{inputenc}
\usepackage{multirow}
\usepackage{float}
\usepackage{bm}
\usepackage{booktabs} 
\usepackage{array} 
\usepackage{graphicx}
\usepackage{amsmath}
\usepackage{amssymb}
\usepackage{color}
\usepackage[per-mode=symbol,separate-uncertainty]{siunitx} 
\usepackage[capitalize]{cleveref} 

\newcommand{\island}{j}
\newcommand{\iisland}{l}

\newcommand{\flux}{\Phi}
\newcommand{\fluxC}{\flux_{\text{C}}}
\newcommand{\fluxT}{\flux_{\text{T}}}
\newcommand{\MagQuan}{\flux_0}
\newcommand{\MagQuanreduced}{\varphi_0}

\DeclareSIUnit{\fluxzero}{\flux_0}

\newcommand{\Ej}{E_{J}}
\newcommand{\Ec}{E_{C}}
\newcommand{\Lj}{L_{J}}
\newcommand{\Ljmin}{L_{J,\text{min}}}
\newcommand{\Cj}{C_{J}}
\newcommand{\Cg}{C_{\text{g}}}
\newcommand{\Nj}{N}
\newcommand{\asymmetry}{d}
\newcommand{\lencell}{a}

\newcommand{\wplasma}{\omega_{\text{p}}}
\newcommand{\Csum}{C_{\Sigma}}

\newcommand{\Eself}{E_{S}}
\newcommand{\EjQ}{E_{J,\text{T}}}
\newcommand{\EcQ}{E_{C,\text{T}}}
\newcommand{\LjQ}{L_{J,\text{T}}}
\newcommand{\CjQ}{C_{J,\text{T}}}
\newcommand{\CshQ}{C_{\text{sh,T}}}
\newcommand{\Csh}{C_{\text{sh}}}
\newcommand{\Cc}{C_{\text{c}}}
\newcommand{\CgQ}{C_{\text{g,T}}}
\newcommand{\CgQs}{C_{\text{g,T2}}}

\newcommand{\CshSquid}{C_{\mathrm{sh,S}}}
\newcommand{\CgSquid}{C_{\mathrm{g,S}}}
\newcommand{\CshEM}{C_{\mathrm{sh,0}}}
\newcommand{\CgQEM}{C_{\mathrm{g,T,0}}}

\newcommand{\Qwidth}{\Gamma_{\text{T}}}
\newcommand{\Qfreq}{\omega_{\text{T}}}
\newcommand{\Qfreqsub}[1]{\omega_{\text{T},#1}}

\newcommand{\Ceff}{C_{\text{eff}}}

\DeclareMathOperator{\nop}{\widehat{n}}
\DeclareMathOperator{\phaseop}{\widehat{\varphi}}

\DeclareMathOperator{\Cm}{\widehat{C}}
\DeclareMathOperator{\Jm}{\widehat{J}}

\newcommand{\wVNA}{\omega_{\text{VNA}}}
\newcommand{\wSMB}{\omega_{\text{MW}}}

\newcommand{\mode}{n}

\newcommand{\LjSonnet}{L_{\text{test}}}
\newcommand{\Zleft}{Z_{\text{left}}}
\newcommand{\Zright}{Z_{\text{right}}}

\begin{document}

\author{Javier Puertas Mart\'inez}
\author{S\'ebastien L\'eger} 
\author{Nicolas Gheeraert}
\author{R\'emy Dassonneville}
\author{Luca Planat}
\author{Farshad Foroughi} 
\author{Yuriy Krupko}
\author{Olivier Buisson} 
\author{C\'ecile Naud}
\author{Wiebke Guichard} 
\author{Serge Florens}
\affiliation{Univ. Grenoble Alpes, CNRS, Grenoble INP, Institut N\'eel, 38000 Grenoble, France}
\author{Izak Snyman} 
\affiliation{Mandelstam Institute for Theoretical Physics, School of Physics, University of the Witwatersrand, 
Johannesburg, South Africa}
\author{Nicolas Roch}
\affiliation{Univ. Grenoble Alpes, CNRS, Grenoble INP, Institut N\'eel, 38000 Grenoble, France}
\email{nicolas.roch@neel.cnrs.fr}

\title{A tunable Josephson platform to explore many-body quantum optics in circuit-QED}

\maketitle

\textbf{
The interaction between light and matter
remains a central topic in modern physics
despite decades of intensive research.
Coupling an isolated emitter to a single mode
of the electromagnetic field is now routinely achieved in the
laboratory~\cite{Haroche:qQ}, and standard quantum optics provides a complete toolbox for
describing such a setup. Current efforts aim to
go further and
explore the coherent 
dynamics of systems containing an emitter coupled to several electromagnetic degrees of freedom 
\cite{Sundaresan:2015eu, Wang:2016bt,Naik:2017bo,Liu:2016ic,Mirhosseini}. Recently, ultrastrong coupling to a transmission line
has been achieved where the emitter resonance broadens to a significant
fraction of its frequency, and hybridizes with a continuum of electromagnetic (EM) modes~\cite{FornDiaz:2016bo,Magazzu:2018is}. In this work we gain significantly improved control over this regime. We do so by combining the simplicity and robustness of a transmon qubit and a bespoke EM environment~\cite{Plourde:2015hh,Rastelli:2017wq,Mirhosseini} with a high density of discrete modes, hosted inside a superconducting metamaterial \cite{Jung:2014jm}. This produces a unique device in which the hybridisation between the qubit and many modes (up to ten in the current device) of its environment can be monitored directly. Moreover the frequency and broadening of the qubit resonance can be tuned independently of each other in situ. We experimentally demonstrate that our device combines this tunability with ultrastrong coupling~\cite{Niemczyk:2010gv,Yoshihara:2016bi,Bosman:2017ep,FornReview} and a qubit nonlinearity comparable to the other relevant energy scales in the system.   
We also develop a quantitative theoretical description that does not contain any phenomenological parameters
and that accurately takes into account vacuum fluctuations of our large scale quantum circuit in the regime of 
ultrastrong coupling and intermediate non-linearity. The demonstration of this new platform combined with a quantitative modelling brings closer the prospect of experimentally studying many-body effects in quantum optics. A limitation of the current device is the intermediate nonlinearity of the qubit. Pushing it further will induce fully developed many-body effects, such as a giant Lamb shift~~\cite{LeHur:2012wq} or nonclassical states of multimode optical fields~~\cite{Goldstein:2013kq,Peropadre:2013iaa,Gheeraert:2017gq,Gheeraert2018}. Observing such effects would establish interesting links between quantum optics and the physics of quantum impurities~\cite{hewson_kondo_1993,LeHur:2016it}. 
}

\noindent\textbf{Introduction}\\
Due to strong interactions between elementary
constituants, correlated solids~\cite{0305-4470-38-8-B01} and trapped cold atoms~\cite{Bloch:2008gl}
host fascinating many-body phenomena.
Attempts to produce similar effects in purely optical systems are hampered by the
obvious fact that photons do not naturally interact with each other. If this obstacle can be overcome,
there is the tantalizing prospect of
probing the many-body problem using the contents of the quantum optics toolbox,
such as single photon sources and detectors, high-order correlations in time-resolved 
measurements, entanglement measures, and phase space tomographies to name a few. 

One route to building a many-body quantum optical system is to rely
on arrays of strongly non-linear cavities or resonators~\cite{Houck:2012iq,Carusotto:2013gh,LeHur:2016it}, 
but minimising disorder in such architectures is a formidable challenge. Another route that circumvents these difficulties involves
coupling a single well-controlled non-linear element to a disorder free harmonic environment. If the difficult experimental challenge of engineering an ultra-strong coupling can be overcome,
thus exceeding the boundaries of the standard single photon regime in quantum optics, this approach could pave the way to bosonic realizations of electronic impurity
systems such as the famous Kondo and Anderson models~\cite{hewson_kondo_1993}.
Our goal here is to achieve
a large coupling between a sufficiently non-linear qubit and a quantum coherent 
environment containing many harmonic degrees of freedom.

When coupling an impurity to a finite size electromagnetic environment, five important frequency scales
have to be considered. The impurity is characterized by its qubit frequency $\omega_\text{qubit}$, i.e. the excitation
frequency between its two lowest internal states.
Real impurities always possess more than two levels. The anharmonicity $\alpha$,
defined as the difference between $\omega_\text{qubit}$ and the frequency 
for excitation from the second to third internal impurity state,
characterises the departure of an impurity 
from a trivial harmonic oscillator ($\alpha\to0$) or a pure two-level system 
($\alpha\to\infty$).
The coupling between
the impurity and environment is characterized by the
spontaneous emission rate $\Gamma$ at which the impurity exchanges energy with its
environment. The environment itself is characterized by its free spectral range $\delta\omega$, which measures the typical
frequency spacing between environmental modes, and the spectral broadening
$\kappa$ of these modes due to their coupling to uncontrolled degrees of
freedom. The sought-after multi-mode regime is obtained when 
$\Gamma$ is larger than $\delta\omega$ so that the impurity is always
coupled to several discrete environmental modes, producing a cluster of
hybridized qubit-environment resonances~\cite{Meiser:2006it}. There are several requirements for reaching the many-body regime. First,
$\Gamma$ must be a significant fraction of $\omega_\text{qubit}$ (ultra-strong coupling). This is a
prerequisite for multiparticle decay~\cite{Goldstein:2013kq,Gheeraert2018}. 
If the coupling is too weak, the system becomes trivial, since only
number-conserving processes are relevant (Equivalently Markov and rotating
wave approximations apply.) A second requirement is $\alpha\gtrsim\Gamma$. 
If this condition is not met, the non-linearity of the impurity is swamped by the broadening of the
impurity levels, and the same frequency will drive transitions between several impurity levels, so that the system as a whole
behaves more like a set of coupled harmonic oscillators than like a two-level system coupled to an environment~\cite{Weissl:2015do}.
Within the many-body regime, two limits can be distinguished. In the case of a finite-size 
environment that we address here (namely, $\delta\omega>\kappa$), each mode of the system 
can be addressed and controlled individually, while in the limit of a thermodynamically 
large environment ($\delta\omega/\kappa\to0$) one recovers a smooth dissipation-broadened 
qubit resonance. 

The many-body ultra-strong coupling regime (defined by the first two conditions
above) is hard to reach in quantum optics experiments because
the coupling to three-dimensional vacuum fluctuations arises at order
$[\alpha_\mathrm{QED}]^3$, with $\alpha_\mathrm{QED}\simeq1/137$ the fine
structure constant~\cite{Haroche:qQ}. However, for superconducting qubits
coupled to transmission lines, the scaling is much more
favorable\cite{Devoret:2007gi,Bourassa:2009gy,Peropadre:2010ie,Peropadre:2013iaa}
than in a vacuum. Indeed the ratio $\Gamma/\omega_\text{qubit}$ can essentially
be made arbitrarily large, provided the impedance of the environment matches
that of the qubit (see Sec. E of the Supplementary Information). Building on this ability of
superconducting circuits to reach very large couplings, several experiments
demonstrated the ultra-strong coupling regime in coupled qubit/cavity 
systems~\cite{Niemczyk:2010gv,FornDiaz:2010by,Yoshihara:2016bi,Bosman:2017ep}.
The rich physics associated to this coupling regime has also been evidenced using quantum simulation~\cite{Braumuller:2016wj,Langford:2017ep}. 
The condition $\Gamma>\delta\omega$ has also been fulfilled by coupling superconducting qubits
to open transmission lines~\cite{Astafiev:2010cm,Hoi:2011tn,Haeberlein:2015um}
or engineered resonators~\cite{Sundaresan:2015eu,Naik:2017bo}. However, it is
only recently that the necessary conditions for the many-body regime were 
demonstrated concurrently~\cite{FornDiaz:2016bo,Magazzu:2018is}. The device
of Refs.~\onlinecite{FornDiaz:2016bo,Magazzu:2018is} consists of a
flux qubit coupled to the continuum provided by a superconducting transmission
line, which realises the thermodynamic limit ($\delta\omega/\kappa\to0$).
Limitations of such setups include the lack of a microscopic model (since it is hard
to characterize the waveguide properties of a transmission line outside the
relatively narrow 4-8 GHz band where microwave transmission experiments can
comfortably be performed), and importantly, that the transmission line is not an
in-situ tunable environment.

\begin{figure*}[htb]
\begin{center}
\includegraphics[scale=0.40]{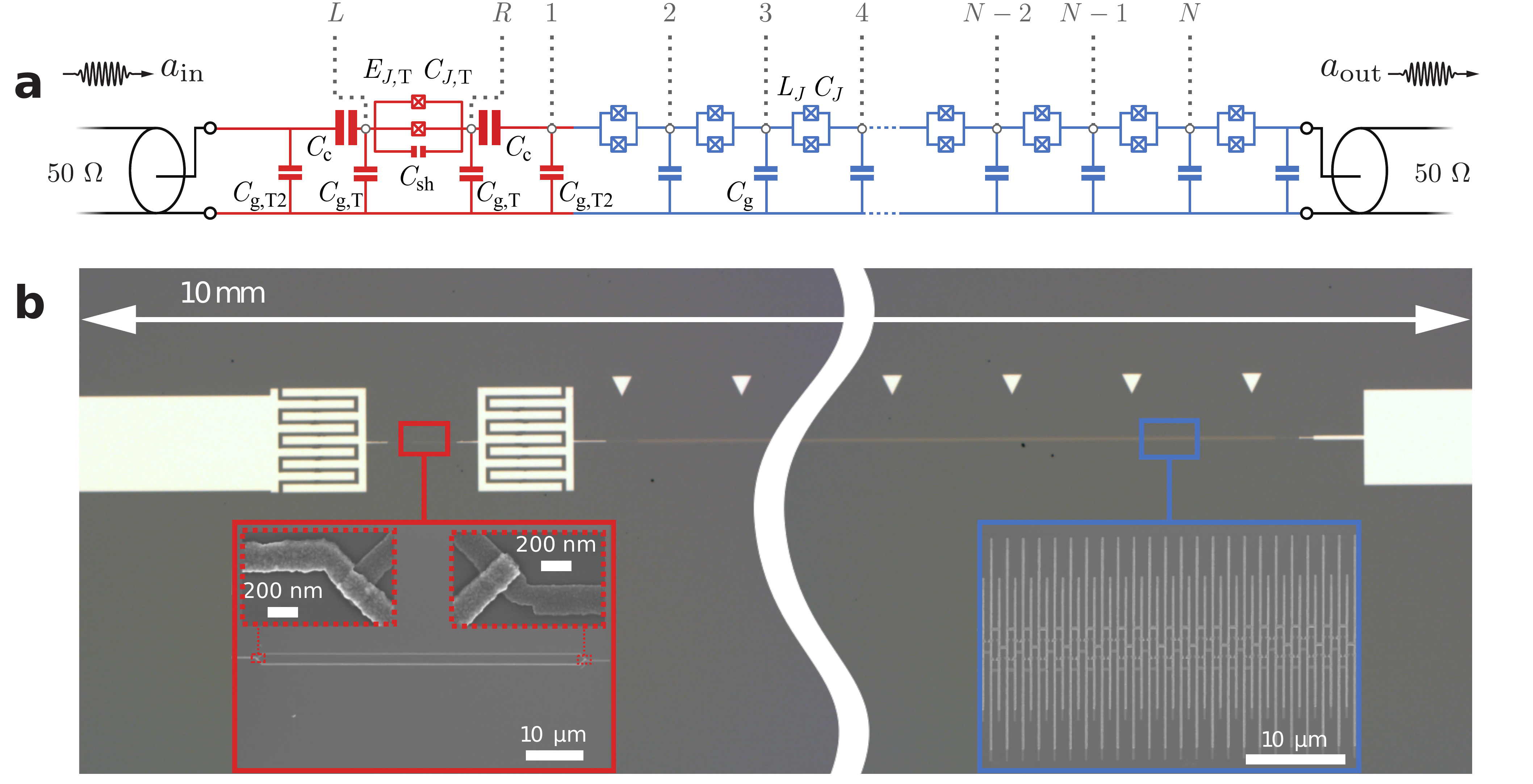}
\caption{\textbf{A Josephson platform for waveguide quantum electrodynamics.} 
\textbf{a} Lumped element model of the circuit including the nodes used in the calculations. 
\textbf{b} Optical
microscope image of the sample. The two zoom-in are SEM pictures of the SQUID of
the qubit (red square) and the SQUIDs in the chain (blue square). The qubit
is capacitively coupled to the chain and to a 50$\Omega$ measurement line, via 
large interdigital contacts. Only a small portion of the Josephson chain, which
comprises 4700 SQUIDs in total, is shown.
}
\label{fig1}
\end{center}
\end{figure*}

\begin{figure*}[htb]
\begin{center}
\includegraphics[scale=0.39]{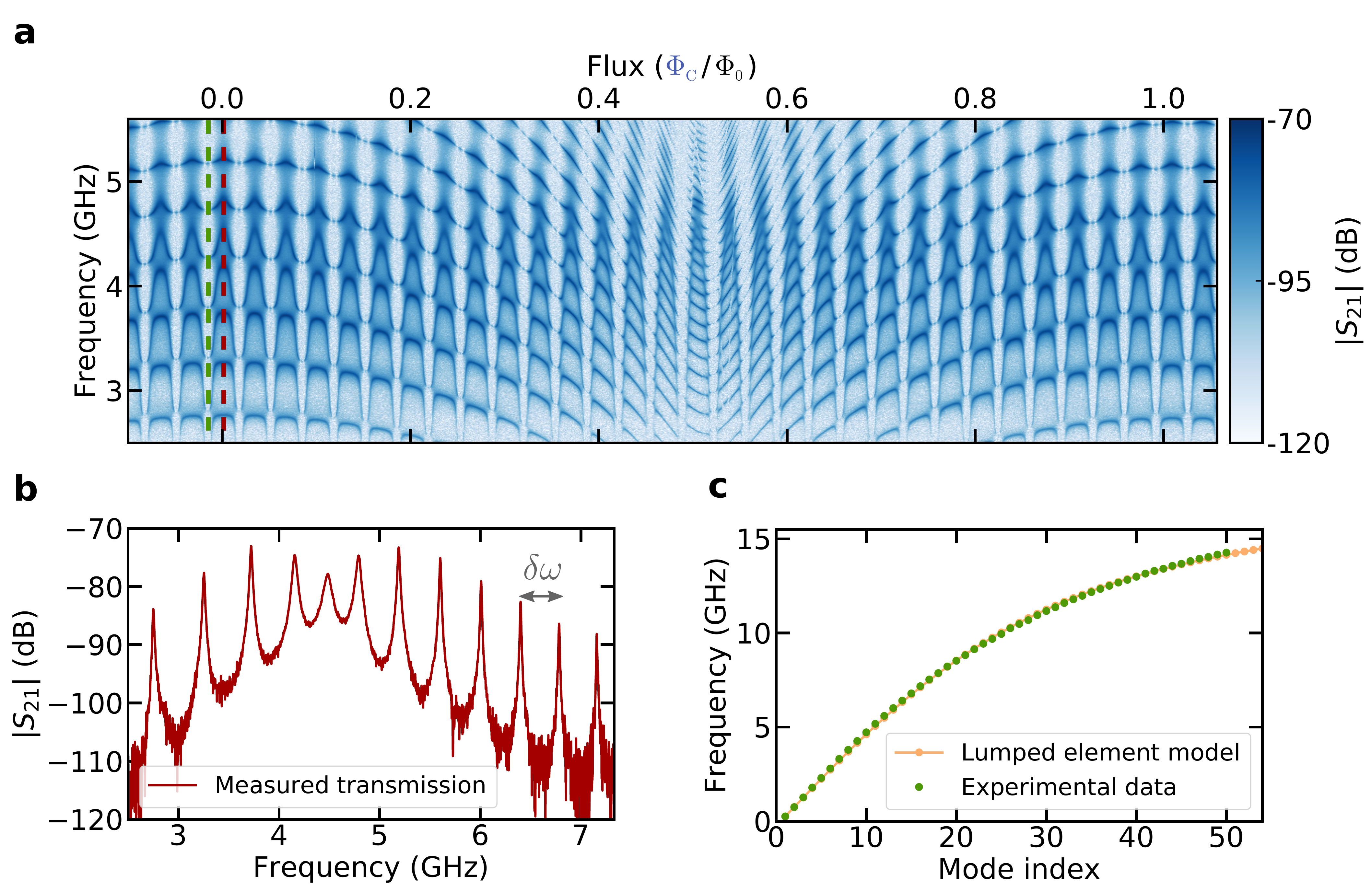}
\caption{\textbf{Spectroscopic analysis of the full quantum circuit.}
\textbf{a} 
Microwave transmission measurement of the complete device (transmon and chain)
shown in panel b of in Fig \ref{fig1}, as a function of flux. 
Two flux periods can be seen, the long one (resp. the short one) being related 
to the SQUIDs in the chain (resp. in the qubit). Two vertical cuts indicate
the spectroscopic traces shown in the two bottom panels respectively. \textbf{b} Frequency
trace of microwave transmission through the device at flux $\fluxC=0$
(red cut in panel a), in which case the transmon flux is also
$\fluxT=0$. The free spectral range $\delta\omega$ is shown in grey. \textbf{c}
Dispersion relation of the chain alone obtained from the fit of the resonances
at chain flux $\fluxC=-0.015\ \MagQuan$ corresponding to $\fluxT=-\MagQuan/2$
(green cut in panel a), so that the chain modes do not hybridize
with the transmon.}
\label{fig2}
\end{center}
\end{figure*}

\noindent\textbf{Results/Discussion}\\
In this work, we circumvent the above limitations, by designing circuits
that provide independent tunability of both a qubit and a finite size but very
large environment, while allowing high-precision spectroscopic measurements of
the environment itself ($\delta\omega>\kappa$) and first principle modeling.
Our device, shown in Figure \ref{fig1}{\bf b}, consists of a transmon qubit,
which is relatively insensitive to both charge and flux noise, capacitively
coupled to a long one-dimensional Josephson metamaterial, comprising 4700 SQUIDs.
Such chains have been studied since the early 90's in the context of the
superconductor-insulator
transition~\cite{Geerligs:1989gx,Chow:1998bq,Fazio:2001jv,Cedergren:2017ef} or
to explore dual of the Josephson effect~\cite{Corlevi:2006vu,Pop:2010bs,Ergul:2013fr}. 
Our setup differs in two ways from these previous works. 
First, we took great care to produce a chain in the linear regime, far from the onset of 
non-linear effects such as quantum phase slips~\cite{Likharev:ik}, so that one of the
basic benifits of quantum optics, i.e. the elimination of non-linearities where they are not wanted, is realized. Second, we performed AC 
microwave spectroscopy of our device, instead of DC transport measurements. This allows us to  
characterize the electromagnetic degrees
of freedom, also called \textit{Mooij-Sch\"on} plasma 
modes~\cite{Mooij:1985bwa,Anonymous:2012jo,Anonymous:2012co,Altimiras:2013et,Weissl:2015do,Muppalla:2018cy},
microscopically. We managed to resolve as many as 50
individual low frequency electromagnetic modes of this non-dissipative and fully tunable environment (See
Figure \ref{fig2}{\bf{c}}). 
An essential property of the Josephson metamaterial is its high characteristic impedance
$Z_c=\sqrt{\Lj/\Cg}\simeq \SI{1590}{\ohm}$, which being of the same order of magnitude as the effective
impedance of our transmon qubit,
$Z_T=\hbar/(2e)^2\sqrt{2\EcQ/\EjQ} \simeq \SI{760}{\ohm}$ (here at zero flux), allow us
to reach multi-mode ultrastrong coupling. The simplicity of the transmon
architecture enables us to either compute from first principles or to extract all the parameters necessary to construct a
microscopic model of the full system, without dropping the so-called
`$A^2$-terms', a routine approximation in
optics\cite{Nataf:2010vl,GarciaRipoll:2015ba,
Malekakhlagh:2016is,Gely:2017cwa,Andersen:be}, that however breaks down at
ultra-strong coupling. 

Our measurements are based on the frequency-resolved microwave transmission through the 
whole device. Figure \ref{fig2}{\bf b}
shows the amplitude of the transmitted field at a fixed value of the external
magnetic field, and at low probe power. This spectrum reveals a set of
resonances in our device, displaying a narrow spectral broadening
$\kappa/2\pi=20$MHz (for the non hybridized modes of the chain) caused
by the coupling to the \SI{50}{\ohm} contacts. As the external magnetic field
is varied, two modulation periods are seen in the resonance spectrum (Figure
\ref{fig2}{\bf a}). The short and long
periods correspond respectively to a one quantum $\MagQuan=h/2e$ increase of the
flux through the large transmon or through the small chain SQUID loops. This feature
allows us to adjust independently the flux threading the transmon SQUID loop
($\fluxT$) from the one threading the chain SQUID loops ($\fluxC$). The former
controls the qubit frequency, while the latter controls the impedance of the
environment, and hence the qubit-environment coupling strength. Before studying the
hybridization between the transmon and chain, we characterize the chain on its
own by setting $\fluxT=\MagQuan/2$, so that the qubit decouples from it (See \cref{fig2}\textbf{c} and Sec. B of the Supplementary Information). We obtain a good fit between the dispersion relation
predicted by a microscopic model and the one extracted from the measured
resonances. This allows us to extract all of the parameters necessary to characterize
the chain modes.
\begin{figure}[htb]
\begin{center}
\includegraphics[scale=0.35]{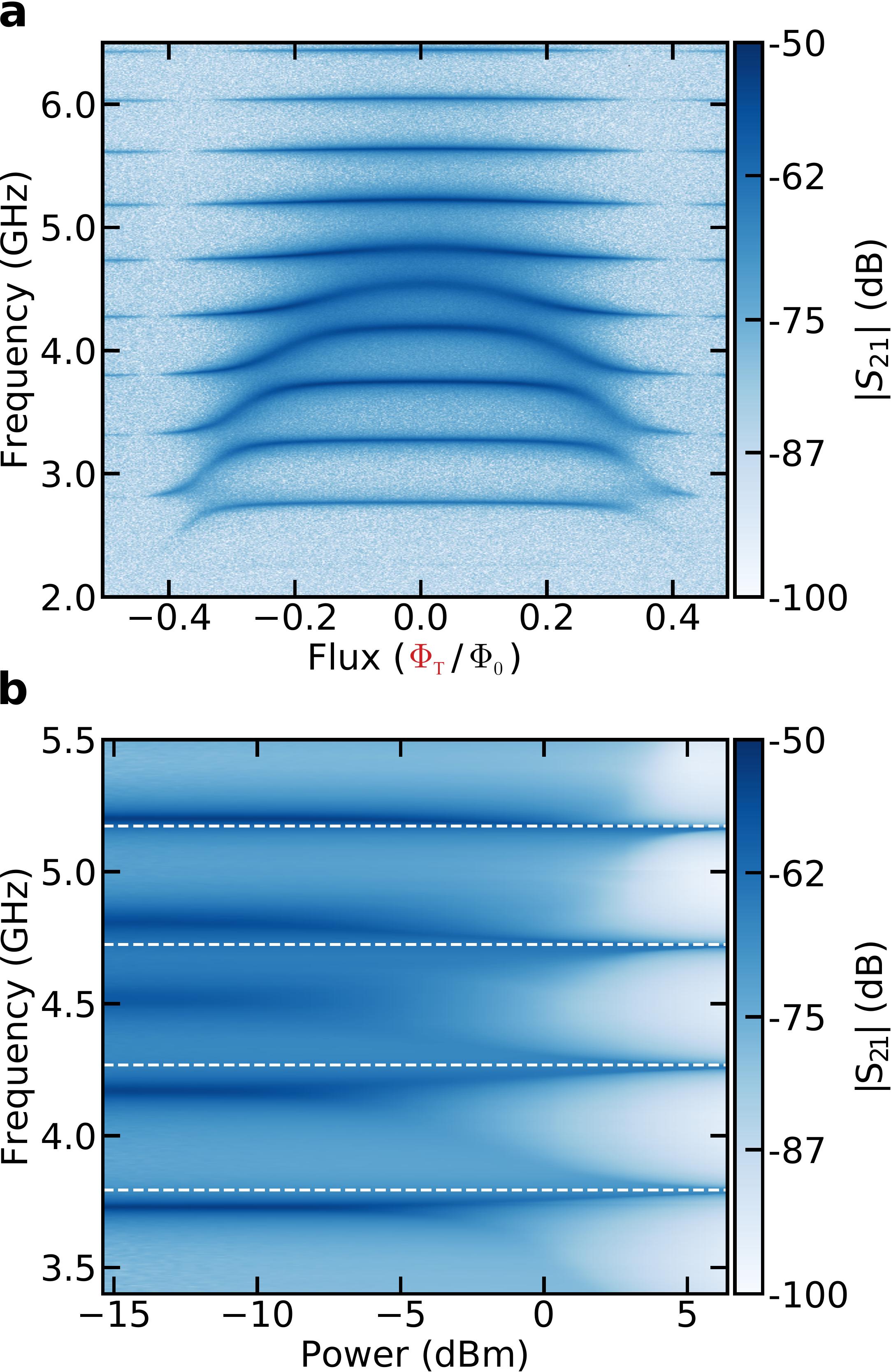}
\caption{
\textbf{Hybridization of qubit and chain modes.}
\textbf{a} Transmission spectrum as a function of transmon flux
$\fluxT$, at chain flux $\fluxC\simeq0$ (representing a small portion
of the full spectroscopy in Fig.~\ref{fig2}{\bf a}). The horizontal lines 
are the chain modes far from the qubit resonance. 
The flux modulation of the transmon frequency produces a bell-shaped 
succession of anticrossings. \textbf{b} Transmission spectrum 
as a function of applied microwave power at fixed fluxes $\fluxT=\fluxC=0$. The white dashed lines indicate the modes of the array at $\fluxT = \MagQuan/2$.
With increasing power, the transmon-like mode near \SI{4.5}{\giga\hertz} disappears, 
showing its non-linear quantum character. In addition the modes of the array shift to their bare frequencies.}
\label{fig3}
\end{center}
\end{figure}

The transmon qubit becomes active when $\fluxT\not=\MagQuan/2$, and is expected
to hybridize with the chain.
\Cref{fig3}\textbf{a} shows a low-power spectroscopy
of the system as a function of $\fluxT$, keeping $\fluxC$ nearly constant. 
When a chain mode is not hybridized with the qubit, the corresponding spectral line runs 
nearly horizontally. When $\fluxT$ is varied, the qubit frequency sweeps
across the resonances of the chain modes and creates a clear pattern of several avoided
level crossings (See \Cref{fig3}\textbf{a}). We note that at fixed $\fluxT$, several 
chain modes in the vicinity of the transmon resonance are visibly displaced.
Thus, the transmon simultaneously hybridizes with many modes. This is a
signature of multi-mode ultra-strong coupling, a topic that will be further addressed 
below. Evidence that the transmon behaves as a qubit is
provided by its saturation spectrum (Figure \ref{fig3}\textbf{b}). 
Here the fluxes $\fluxT$ and $\fluxC$ are kept constant, while the transmission through 
the system is recorded at increasing probe power. For a harmonic system, the resonance positions are independent of driving power.
In a anharmonically oscillating classical system, a gradual dependence on the driving
power would appear. Experimentally, we observe that
below a driving power $\sim -10$ dBm, the resonance
positions in the transmission spectrum are independent of the driving probe
power. As the driving power increases beyond  $-10$ dBm, the peak 
around \SI{4.6}{\giga\hertz} disappears while the other peaks assume the
positions they have when the qubit is inactive. (The horizontal lines correspond to the
low power spectrum when $\fluxT =\MagQuan/2$.)  
This is evidence of saturation,
a clear qubit signature~\cite{Astafiev:2010cm,Hoi:2011tn, LeHur:2012wq}.
The fact that several chain modes are shifted when the transmon
is saturated constitutes an additional proof that the transmon hybridizes with 
many modes at once.

\begin{figure}[htb]
\begin{center}
\includegraphics[scale=0.35]{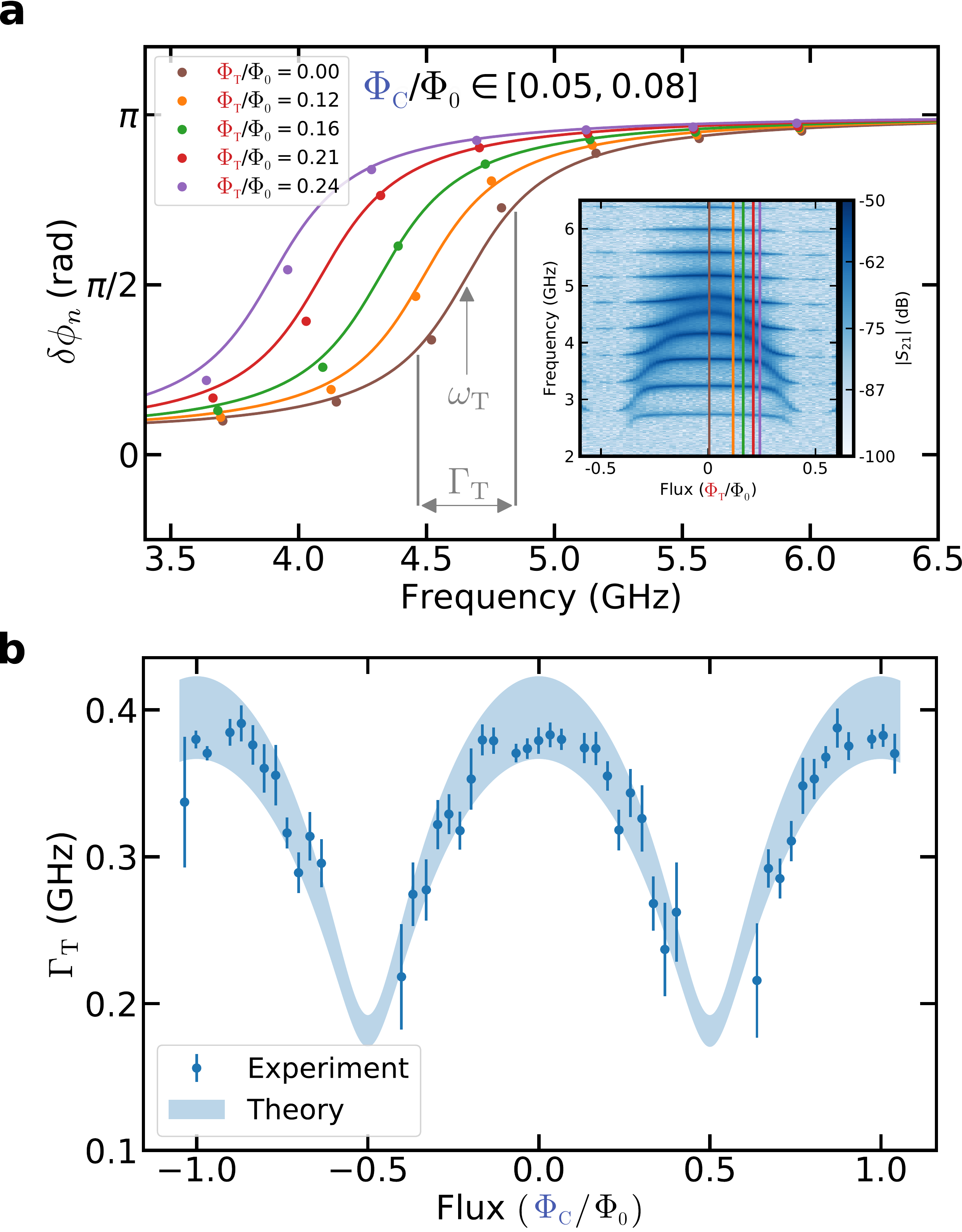}
\caption{
\textbf{Extraction of qubit properties from the measurement of its controlled
environment}
\textbf{a} Phase shift $\delta\phi_n$ of the discrete chain modes as
a function of mode frequency $\omega_n$ for different transmon fluxes $\fluxT$,
fixing $\fluxC=0$. The solid line is a 
fit using \cref{deltaphi}. 
The inset shows the chosen transmon fluxes $\fluxT$ as line cuts in the 
transmission measurement (with the same color code).
\textbf{b} The transmon width $\Qwidth$ for different fluxes in 
the chain $\fluxC$, showing control of the coupling to the large but discrete
environment. The experimental points (dots) are obtained from an arctangent fit of the
data in panel a. For better
visibility, only the flux values where the transmon frequency is maximum are
included. The blue shaded area represents the theoretical
expectation for $\Qwidth$, within a confidence interval given by the error in
the capacitances in \cref{tab:sample_parameters}.}
\label{fig4}
\end{center}
\end{figure}

To quantify the hybridization of the transmon mode with the chain modes, we
compare the normal mode spectrum of the full system at $\fluxT\neq\MagQuan/2$ to
the spectrum at $\fluxT=\MagQuan/2$. As mentioned above, in the latter case, the
transmon decouples from the bare modes of the chain. The
system with $\fluxT\neq\MagQuan/2$ therefore has one extra mode in the vicinity of
the transmon frequency.
We define the relative frequency shift $\delta\phi_n(\fluxT,\fluxC)$ as the
difference in frequency between the $n$th mode of the coupled and uncoupled
chains, normalized
to the free spectral range of the chain $\delta\omega_n(\fluxT,\fluxC) = 
\omega_n(\MagQuan/2,\fluxC)-\omega_{n-1}(\MagQuan/2,\fluxC)$, and including a $\pi$ factor
for later convenience, i.e.
\begin{equation}
\delta\phi_n(\fluxT,\fluxC)=\pi \frac{\omega_n(\MagQuan/2,\fluxC)-\omega_n(\fluxT,\fluxC)}
{\delta\omega_n(\fluxT,\fluxC)},\label{df1}
\end{equation}
where $\omega_n(\fluxT,\fluxC)$ is the frequency of the $n$th lowest non-zero mode 
for a given flux in the transmon and in the chain. This frequency shift is
readily extracted from the peak positions in our global spectroscopic map
(Fig.~\ref{fig2}{\bf a}). Remarkably (see Sec. H of the Supplementary Information for a derivation), 
$\delta\phi_n(\fluxT,\fluxC)$ in Eq.~(\ref{df1}) equals the phase shift
experienced by mode $n$ due to 
the presence of the nearby transmon mode:
\begin{equation}
\delta\phi_n(\fluxT,\fluxC)=\phi_n(\fluxT,\fluxC)-\phi_n(\MagQuan/2,\fluxC),
\label{deltaphi}
\end{equation}
where $\phi_n(\fluxT,\fluxC)$ is the phase shift of mode $n$ of the full system at transmon flux $\fluxT$
and chain flux $\fluxC$. From Eq.~(\ref{df1}) it follows that the phase shift equals 
0 (resp. $\pi$) for modes far below (resp. far above) the renormalized transmon
frequency. For hybridized modes in the vicinity of the transmon line, $\delta\phi_n(\fluxT,\fluxC)$ 
lies between 0 and $\pi$. This behavior is clearly observed in \cref{fig4}\textbf{a} where the 
measured relative frequency shifts are 
reported for a chain flux $\fluxC = 0$ and various transmon fluxes
$\fluxT$. The wide frequency dispersion of intermediate $\delta\phi_n(\fluxT,\fluxC)$ provides 
direct evidence for a hybridization with up to ten chain modes. In the thermodynamic
limit of an infinite chain with perfect impedance matching to the
measurement ports, the transmon-induced phase shift $\delta\phi_n(\fluxT,\fluxC)$ becomes a continuous function 
$\delta\phi(\omega,\fluxT,\fluxC)$ of
the mode frequency $\omega$.
Moreover, it can be shown that the frequency derivative of
$\delta\phi(\omega,\fluxT,\fluxC)$ matches very precisely the theoretically
expected lineshape of the dissipative response of the transmon coupled to an
infinite environment. (See Sec. I of the Supplementary Information.) This constitutes a
central finding of our work: the renormalized 
transmon frequency $\Qfreq$ and linewidth $\Qwidth$ can be directly inferred from 
a measurement of the phase shifts of the individual modes in the finite bath.
In terms of measurement protocol however, there is a sharp difference
between the chain mode phase shifts and the qubit response functions. 
Usually, the qubit response is obtained by observing the transmon, and its environment 
can be viewed as a black box that combines unmonitored decoherence channels as well 
as the physical ports used for measurement. This procedure constitutes the usual paradigm 
in the study of open quantum systems. Our protocol is unusual because information about 
an open quantum system is obtained by monitoring the discrete modes that
constitute its dominant environment.

We finally turn to a quantitative analysis of our data, including a comparison to 
the predictions of a microscopic model, in order to determine if the requirements for
reaching the many-body regime has been met. By extracting 
the maximum renormalized transmon frequency $\omega_\mathrm{T,max}$  
extracted from the phase shift data of~\cref{fig4}\textbf{a} at chain flux $\fluxT$ an integer multiples of $\flux_0$, we are able to 
infer the only remaining unknown system parameter, namely the maximum transmon Josehpson energy $E_{J,\text{T,max}}$.
This allows us to estimate the anharmonicity of the transmon $\alpha$,
which ranges from $\SI{0.36}{\giga\hertz}$ at $\flux_\text{T}=0$ to $\SI{0.44}{\giga\hertz}$ at $\flux_\text{T}=0.3\flux_0$.  
We emphasize that the condition
$\alpha\gtrsim\Qwidth$ for anharmonic many-body behavior is thus fulfilled 
(see~\cref{fig4}\textbf{b} for the extracted transmon linewidth $\Qwidth$,
which lies in the range 0.2-0.4GHz). 
Using the extracted parameters to
calculate $\delta\phi(\omega,\fluxT,\fluxC)$ according to
Eq.~(\ref{eq:phi_analytic}) we find the predicted theoretical lines
in \cref{fig4}\textbf{a}. 
The excellent agreement between theory and experiment seen here for six different
values of $\fluxT$ persists for each of the hundreds of ($\fluxT$,$\fluxC$) combinations
where we have made the comparison. (See Sec. C of the Supplementary Material for a further
selection of results.) We stress that this agreement 
is obtained after all model parameters have been fixed, so that there is no fitting involved
in comparing the predicted and measured phase shifts. The quantitative modeling of such a
large quantum circuit clearly 
is an important landmark 
in the field of open quantum systems. In \cref{fig4}\textbf{b} we examine the
transmon linewidth $\Qwidth$ that we extracted from the phase shift data, as a
function of chain flux $\fluxC$ for fixed transmon flux $\fluxT=\num{0}$. Very good
agreement (with no fitting
parameters) is again obtained with the prediction of our model. 
These results demonstrate that we can tune the qubit-environment
coupling independently from $\Qfreq$ using the flux in the chain, and that we
achieved the ultra-strong coupling in our waveguide, \textit{i.e.} coupling to a 
large number (here 10) of modes with a sizeable linewidth $\Qwidth/\Qfreq\simeq0.1$.
A hallmark of ultra-strong coupling is the failure of the rotating wave
approximation (RWA), as previously discussed in coupled qubit and cavity 
systems~\cite{Niemczyk:2010gv}. We have examined the consequences of the RWA on our microscopic model (see Sec. J of the Supplementary 
Information), and found a discrepancy of 100MHz in the transmon frequency~$\omega_\mathrm{T}$, showing the quantitative importance of non-RWA terms. We would like to stress that demonstrating the relevance of these terms is much more than obtaining a good data-theory agreement. With counter-rotating contributions in the few percent range, we expect a finite rate for parametric processes in which photon-number is not conserved. In future we plan to use the current platform to observe these interesting many-body effects directly. 

In conclusion, this work provides the first demonstration of many-body ultra-strong 
coupling between a transmon qubit and a large and tunable bath. To obtain full control
over the environment, a superconducting metamaterial, comprising 4700 SQUIDs, 
was employed. Although this quantum circuit contains a huge number 
of degrees of freedom, we were able to characterise all its parameters in-situ.
This allows us to demonstrate unambiguously that our systems meets the three
conditions required to reach the many-body regime, namely $\Gamma\gtrsim\delta\omega$,
$\Gamma\gtrsim0.1\omega_\text{qubit}$ and $\alpha\gtrsim\Gamma$.
A novel experimental methodology was implemented to analyze the qubit
properties by means of the extraction of the phase shifts of the 
environmental modes. 
Despite the large size of our quantum circuit, we succeeded
in providing a fully microscopic model which accounts for the transmon response
without any fitting parameters. We also found that the qubit linewidth for the
long chain agreed with results in the thermodynamic limit, showing that the finite
environment has the same influence on the qubit as a truly macroscopic bath.
The further possibility to tune the coupling to the environment in-situ,
demonstrated by a 50\% flux-modulation of the qubit linewidth, opens the way to
controlled quantum optics experiments where many-body effects are
fully-developped\cite{LeHur:2012wq,Goldstein:2013kq,SanchezBurillo:2014ia,Snyman:2015dm,Gheeraert:2017gq,Gheeraert2018},
as well as more advanced environmental engineering for superconducting
qubits\cite{Houck:2012iq}.\\

\noindent\textbf{Methods}\\
\footnotesize
\noindent{\bf Sample fabrication and parameters}\\
\noindent
The sample was fabricated on a highly resistive silicon substrate, using a
microstrip geometry. The ground is defined as the backside of the wafer which was
gold-plated, ensuring a good electrical conductivity. Interdigital
capacitances were chosen to connect the transmon and the metamaterial. They
do not provide the lowest surface participation factor\cite{Wang:2015tp,
Dial:2015wk} but they allow us to maximize the coupling capacitances $\Cc{}$ 
of the transmon to the chain, while minimizing the capacitances to ground 
$\CgQ{}$ and $\CgQs{}$. 
(See panel \textbf{a} of \cref{fig1} for the definitions of the these capacitances.)
This
system is probed via two \SI{50}{\ohm} transmission lines, one of which is 
capacitively coupled to the transmon, while the other is galvanically
coupled to the chain. The whole device (Josephson junctions, capacitances
and transmission lines) was fabricated in a single electron-beam lithography
step, using a bridge-free fabrication technique\cite{Lecocq:2011dk}. The
Josephson elements of the chain are tailored to be deep in the linear regime 
($\Ej{}/\Ec{}$ = 8400), where $\Ej$ and $\Ec$ are the respectively the Josephson energy at $\fluxC=0$
and the charging energy of a chain element), leaving the transmon as the main source of non-linearity 
in the system. All parameters of the system are listed in \cref{tab:sample_parameters}. 
Here, $\Ljmin$ is
the minimum inductance of a chain SQUID loop, (which occurs when $\fluxC=0$). There is a
slight asymmetry between the two Josephson junctions that constitute a single chain SQUID loop, which is
quantified by the asymmetry parameter $d$. The flux-dependent inductance $\Lj(\fluxC)$ of a chain SQUID loop
is given by
\begin{equation}
\Lj(\fluxC)=\frac{\Ljmin}{\sqrt{\cos^2\left(\pi\fluxC/\flux_0\right)+d^2\sin^2\left(\pi\fluxC/\flux_0\right)}},
\end{equation}
and the corresponding Josephson energy by $\Ej(\fluxC)=\MagQuanreduced^2/\Lj(\fluxC)$ with 
 $\MagQuanreduced = \hbar/2e$ the reduced flux quantum. The transmon Josephson junctions are symmetric,
 and hence the flux dependent transmon Josephson energy is given by
 \begin{equation}
\EjQ(\fluxT)=E_{J,\text{T,max}}\left|\cos\left(\pi \fluxT/\flux_0\right)\right|\label{ejtdef}
\end{equation}
The transmon charging energy is not an independent parameter (see \cref{eq:charging_energy}), 
but is listed in the table, due to the prominent role it plays in what follows. The meanings of 
the remaining parameters in \cref{tab:sample_parameters} are explained in panel \textbf{a} of \cref{fig1}. 
The majority of the parameter values listed are obtained either using a finite-element solver (Sonnet) or 
extracted from the measured dispersion relation of the chain 
at $\fluxT=0$. (See Sec. B and D of the suplementary Material.) The only exceptions are $\Cj$, that is obtained from knowledge of the junction areas via an empirical formula, 
and $E_{J,\text{T,max}}$ that we extract via a procedure described in
the last subsection below, which uses the data at $\Phi_T$ equal to multiples of $\Phi_0$  in panel \textbf{a} of Fig. \ref{fig2}.
\begin{table}[htbp]
\footnotesize
\hspace{-1.9cm}
	\parbox{.30\linewidth}{
		\begin{tabular}{l r}
		\hline
		\multicolumn{2}{c}{\textbf{Chain parameters}} \\
		$\Ljmin$ & \SI{0.33 \pm 0.02}{\nano\henry} \\
		$\Cg$ & \SI{0.13 \pm 0.01}{\femto\farad} \\
		$\Cj$ & \SI{259 \pm 14}{\femto\farad} \\
		$\Nj$           & \num{4700}                    \\
		$\asymmetry$ (asymmetry)            & \num{0.25}                    \\
		\hline
		\end{tabular}
	}
\hspace{1.7cm}
	\parbox{.30\linewidth}{
		\begin{tabular}{l r}
		\hline 
		\multicolumn{2}{c}{\textbf{Transmon qubit parameters}} \\
		$\CgQs$   & \SI{33 \pm 1}{\femto\farad} \\
		$\CgQ$  & \SI{48 \pm 2}{\femto\farad} \\
		$\Cc$  & \SI{119 \pm 2}{\femto\farad} \\
		$\Csh$  & \SI{6.9 \pm 0.1}{\femto\farad} \\
		$\CjQ$  & \SI{5.2 \pm 0.3}{\femto\farad} \\
		$E_{J,\text{T,max}}/h$   & \SI{10.2 \pm 0.4}{\giga\hertz} \\
		$\EcQ/h$ & \SI{2.4 \pm 0.1}{\giga\hertz} \\
		\hline
		\end{tabular}
	}
\caption{Sample parameters}
\label{tab:sample_parameters}
\end{table}

\noindent{\bf Full Model}\\
\noindent
The circuit diagram for the lumped-element model is shown in \cref{fig1}. It
consists of $\Nj+2$ nodes, where $\Nj$ is the number of SQUIDs in the chain. To
describe the circuit, we use the Cooper pair number operator $\nop_{\island}$,
which gives the number of Cooper pairs in node $\island$, and the
superconducting phase operator $\phaseop_{\island}$, which gives the
superconducting phase at node $\island$. They satisfy the canonical commutation
relations $[\nop_{\island},\phaseop_{\iisland}]=i\delta_{\island
\iisland}$\cite{Devoret1995}. Here $\island, \iisland \in
[L,R,1,2,3,\ldots,\Nj]$ with $L$ and $R$ referring to the left and right
transmon nodes. As explained before, the SQUIDs of the chain are linear
inductors, to a very good approximation. We define $\vec{\nop}^T =
(\nop_L,\nop_R,\nop_1,\ldots,\nop_{\Nj})$ and $\vec{\phaseop}^T =
(\phaseop_L,\phaseop_R,\phaseop_1,\ldots,\phaseop_{\Nj})$. In this notation, the
Hamiltonian of the circuit is given by
\begin{equation}
H = \frac{(2e)^2}{2}\vec{\nop}^T\Cm^{-1}\vec{\nop} - \frac{1}{2}\vec{\phaseop}^T\Jm\vec{\phaseop} 
- \EjQ(\fluxT)\cos\left(\phaseop_R - \phaseop_L\right)
\label{fullh}
\end{equation}
$\Cm$ is the capacitance
matrix, such that elements $[\Cm]_{jl}=[\Cm]_{lj}$ equal the capacitive coupling
between the charges on islands $\island$ and $\iisland$. In the same way,
elements $[\Jm]_{jl}=[\Jm]_{lj}$ of matrix $\Jm$ contains the Josephson energy
that couples the superconducting phase on island $\island$ and island
$\iisland$. Both matrices are $(\Nj+ 2) \times (\Nj + 2)$. Their explicit forms are
given below in \cref{eq:capacitance_matrix_real,eq:J_matrix_real}. In both
matrices, the boundary conditions that determine entries $(L,L)$ and $(N,N)$ are
obtained by assuming that the nodes to the left of node $L$ and to the right of
node $N$ are grounded.
\begin{equation}
\Cm = 
	\begin{pmatrix}
		C_{0} & -\CshQ     &0         &0         &0          &0            &\cdots&0\\
		   -\CshQ &   C_{0}  &-\Cc      &0         &0          &0            &\cdots&0\\
		      0  &-\Cc       &C_{1}&-\Cj      &0          &0            &\cdots&0\\
			0      &0          &-\Cj       &\Csum&-\Cj        &0            &\cdots&0\\
			\vdots &\vdots     &\vdots    &\ddots    &\ddots     &\ddots       &\cdots&0\\
			0      &0          &0         &0      &-\Cj  &\Csum        &-\Cj      &0\\
			0      &0          &0         &0         &0      &-\Cj    &\Csum  &-\Cj \\
			0      &0          &0         &0         &0          &0            &-\Cj   &\Csum
	\end{pmatrix}
	\label{eq:capacitance_matrix_real}
\end{equation}
The elements in the capacitance matrix are given by
\begin{align}
C_0 &= \Cc + \CshQ + \CgQ \nonumber\\
C_1 &= \Cc + \Cj + \CgQs \nonumber\\
\Csum &= 2\Cj + \Cg  \nonumber \\
\CshQ &= \CjQ + \Csh  \nonumber 
\end{align}
\begin{equation}
\Jm = \frac{\MagQuanreduced^2}{\Lj\left(\fluxC\right)}
	\begin{pmatrix}
		0  &  0   &0         &0         &0          &0            &\cdots&0\\
		   0 &   0  &0      &0         &0          &0            &\cdots&0\\
		      0  &0       &1&-1&0          &0            &\cdots&0\\
			0      &0          &-1       &2&-1      &0            &\cdots&0\\
			\vdots &\vdots     &\vdots    &\ddots    &\ddots     &\ddots       &\cdots&0\\
			0      &0          &0         &0      &-1  &2       &-1    &0\\
			0      &0          &0         &0         &0      &-1  &2  &-1 \\
			0      &0          &0         &0         &0          &0            &-1  &2
	\end{pmatrix}
	\label{eq:J_matrix_real}
\end{equation}
We define the operators $\nop_\text{T} = (\nop_R-\nop_L)/2+\,\mbox{constant}$
and $\phaseop_\text{T} = (\phaseop_R - \phaseop_L)/2$ associated with the
transmon dynamics. Introducing these operators, and noting that the total
transmon charge $\nop_R+\nop_L$ is concerved,
 we can rewrite the Hamiltonian as
\begin{align}
H =& \frac{\EcQ{}}{2}\nop_\text{T}^2-\EjQ(\fluxT){}\cos(\phaseop_\text{T})+\frac{(2e)^2}{2}\sum_{\island\iisland=1}^{\Nj}\nop_\island\left[\Cm^{-1}\right]_{\island,\iisland}\nop_{\iisland}+\nonumber\\
&+\frac{\MagQuanreduced^2}{2\Lj\left(\fluxC\right)}\sum_{\island=1}^{\Nj}\left(\phaseop_{\island+1}-\phaseop_{\island}\right)^2+\nop_\text{T}\sum_{\island=1}^\Nj\nu_\island\nop_\island
\label{hamiltonian}
\end{align}
where we defined $\phaseop_{\Nj+1}\equiv 0$. The transmon charging energy $\EcQ$ is given by
\begin{equation}
\EcQ=(2e)^2\left\{\left[\Cm^{-1}\right]_{LL}+\left[\Cm^{-1}\right]_{RR}-2\left[\Cm^{-1}\right]_{LR}\right\}.
\label{eq:charging_energy}
\end{equation}
The coupling of $\nop_\text{T}$ to the charge on island $\island$ is given by 
\begin{equation}
\nu_\island=(2e)^2\left\{\left[\Cm^{-1}\right]_{R\island}-\left[\Cm^{-1}\right]_{L\island}\right\}.
\label{eq:coupling}
\end{equation}

\noindent{\bf Chain modes phase shift in the thermodynamic limit }\\
\noindent
In Fig. \ref{fig4}\textbf{a} we compare the measured relative frequency shift
$\delta\phi_n(\fluxT,\fluxC)$ to the theoretically predicted transmon phase
shift $\delta\phi(\omega,\fluxT,\fluxC)$ with which it is expected to agree in
the thermodynamic limit. Here we provide the analytical formula for the phase
shift $\phi(\omega,\fluxT,\fluxC)$ of a mode with frequency $\omega$. (See Sec. G of the Supplementary
Information for the derivation.) It reads
\begin{equation}
\tan{\phi(\omega,\fluxT,\fluxC)} =
\frac{\Cg-2\Ceff(\fluxT,\omega)}{\sqrt{\Cg(\Cg+4\Cj)}}\frac{1}{\sqrt{\left(\frac{\wplasma(\fluxC)}{\omega}\right)^2-1}}.
\label{eq:phi_analytic}
\end{equation}
In this expression $\wplasma(\flux_C)= 1/\sqrt{\Lj(\fluxC)\left(\Cj+\Cg/4\right)}$ is the plasma frequency of
the chain, and
\begin{equation}
\Ceff(\fluxT,\omega) = C_1 - \Cj -
\frac{\frac{\Cc^2}{C_0-\CshQ}\left[\frac{(\hbar\omega)^2}{(2e)^2} C_0 -
\Eself(\fluxT)\right]}{(C_0+\CshQ)\frac{(\hbar\omega)^2}{(2e)^2} 
- 2\Eself(\fluxT)}.
\end{equation}
has dimensions of capacitance. Finally, $\Eself(\fluxT)$ is an effective linear inductor energy associated with the Josephson junctions in the transmon, which nonetheless incorporates the transmon non-linearity, and
is given by
\begin{equation}
\Eself(\fluxT)=\EjQ(\fluxT)-\sqrt{\EjQ(\fluxT)\EcQ}/4.\label{eq:eself}
\end{equation}
(See Sec. F of the Supplementary Material for further detail.) 
The theoretical $\delta\phi(\omega,\fluxT,\fluxC)$ curves plotted in panel \textbf{a} of \cref{fig4}
were obtained from $\phi(\omega,\fluxT,\fluxC)$ similarly to \cref{deltaphi} as the difference
\begin{equation}
\delta\phi(\omega,\fluxT,\fluxC)=\phi(\omega,\fluxT,\fluxC)-\phi(\omega,\flux_0/2,\fluxC).\label{eq:relshift}
\end{equation}

\noindent{\bf Analysis of the experimental data}\\
\noindent
To extract the relative frequency shift $\delta\phi_n(\fluxT,\fluxC)$ from the
data presented in \cref{fig2} of the main text, we go about as follows. At a
fixed value of the magnetic field that determines $\fluxT$ and $\fluxC$, we fit
each of the peaks in the transmission spectrum individually with a Lorentzian.
This gives the center frequency of the peaks. From these peak positions, we
obtain $\delta\phi_n(\fluxT,\fluxC)$ experimentally using \cref{df1} at a
particular $\fluxT$ and $\fluxC$.
Next we extract the transmon frequency $\Qfreq$ for the transmon coupled to the
chain from the experimentally determined $\delta\phi_n(\fluxT,\fluxC)$. The details are as follows. 
Empirically, we find that the experimentally determined $\delta\phi_n(\fluxT,\fluxC)$ vs. $\omega_n$ data-points
fit an arctangent lineshape
\begin{equation}
F(\omega) = (1-A)\left(\frac{1}{\pi}\arctan{\left(\frac{2\left(\omega-\Qfreq\right)}{\Qwidth}\right)} + \frac{1}{2}\right) + A
\label{eq:arctan_fit}
\end{equation}
very well, for suitable choices of the parameters $A$, $\Qfreq$, and $\Qwidth$.
(In the parameter regime where our device operates, the theoretically predicted phase shift $\delta\phi(\omega,\fluxT,\fluxC)$ 
also closely approximates this line shape.) 
We therefore fit the measured $\delta\phi_n$ vs. $\omega_n$ at fixed $\fluxT$ and $\fluxC$ to \cref{eq:arctan_fit},
interpreting $\Qfreq$ as the frequency and $\Qwidth$ as the resonance width of the transmon when it is coupled to the chain. 
Before we can quantitatively compare the experimental results for $\delta \phi_n(\fluxT,\fluxC)$ to the theoretically predicted
$\delta \phi(\omega,\fluxT,\fluxC)$ (\cref{eq:phi_analytic}), one final model
parameter, namely the maximum transmon Josephson energy $E_{J,\text{T,max}}$ must be
determined from the experimental data. The general procedure is as follows.
Our theoretical model predicts that in the
regime where the actual device operates, this transmon frequency is very nearly
equal to the isolated ($\Lj\to\infty$) transmon frequency, i.e. the chain only
slightly renormalizes the transmon frequency, and indeed, we see little $\fluxC$
dependence in the extracted $\Qfreq$. At $\fluxT=n\flux_0$, $n=0,\,\pm1,\,\pm2,\,\ldots$, where
the transmon Josephson energy is maximal, we therefore use the isolated transmon result 
 (see Sec. F of the Supplementary Information):
\begin{equation}
\Qfreq(\fluxT=n\flux_0)=\sqrt{\EcQ E_{J,\text{T,max}}}-\EcQ /8
\end{equation} 
Taking the average over $n$ of the experimentally determined $\Qfreq(\fluxT=n\flux_0)$, and using our first principle estimate
for $\EcQ$ in Table \ref{tab:sample_parameters}, we obtain
\begin{align}
\Qfreqsub{\text{max}}/2\pi &\equiv \Qfreq(\fluxT=0)/2\pi = \SI{4.64 \pm 0.01}{\giga\hertz},\\
E_{J,\text{T,max}}/h &= \SI{10.2 \pm 0.4}{\giga\hertz} 
\label{eq:wTmax_average_deviation}
\end{align}
The theoretical curves in \cref{fig4}\textbf{a} were then obtained using 
the system parameters in Table \ref{tab:sample_parameters} in
Eqs. (\ref{eq:phi_analytic}), (\ref{eq:relshift}), (\ref{eq:eself}), and (\ref{ejtdef}).
The full data set covers many transmon periods. Within a given transmon
period, we generally analyze data at several values of $\fluxT$ in the interval
from -0.3 $\phi_0$ to 0.3 $\phi_0$. Each transmon period is measured at different $\fluxC$. 
We also take into account the small variation in $\fluxC$ as the transmon flux sweeps through one flux quantum. 
The experimental points in \cref{fig4}\textbf{b} are obtained as the transmon
width $\Qwidth$ closest to $\fluxT = 0$. The error bars come from the least
square fit using \cref{eq:arctan_fit}. The theoretical width is obtained from a
fit of the phase shift $\delta\phi(\omega,\fluxT,\fluxC)$ with the
arctangent of \cref{eq:arctan_fit}.\\

\normalsize
\noindent{\bf Data Availability}\\
\footnotesize
\noindent
The data that support the findings of this study are available from the corresponding
author upon reasonable request.\\

\normalsize
\noindent{\bf Acknowledgements}\\
\footnotesize
\noindent
The authors would like to thank F. Balestro, D. Dufeu, E. Eyraud, J. Jarreau, J. P. Girard, T. Meunier and W. Wernsdorfer, for early support with
the experimental setup. Very
fruitful discussions with C. K. Andersen, H. U. Baranger, D. Basko, S. Bera, J. J. Garcia
Ripoll, S. M. Girvin, F. Hekking, K. Le Hur, C. Mueller, A. Parra and E. Solano are strongly
acknowledged. The sample was fabricated in the clean rooms Nanofab and PTA (Upstream Technological Platform).This research was supported by the ANR under contracts CLOUD
(project number ANR-16-CE24-0005), GEARED (project number ANR-14-CE26-0018),
by the UGA AGIR program, by the National Research Foundation of South Africa 
(Grant No. 90657), and by the PICS contract FERMICATS. J.P.M. acknowledges
support from the Laboratoire d\textquoteright excellence LANEF in Grenoble
(ANR-10-LABX-51-01). R.D. and S.L. acknowledge support from the CFM
foundation.\\

\normalsize
\noindent{\bf Competing interests}\\
\footnotesize
\noindent
The authors declare no competing financial or non-financial interests.\\

\normalsize
\noindent{\bf Author contributions}\\
\footnotesize
\noindent
J.P.M., S.F. and N.R designed the experiment. J.P.M. and N.R. fabricated the 
device. J.P.M. and S.L. performed the experiment and analysed the data with 
help from S.F, N.R. and I.S., while S.F. and I.S. provided the theoretical 
support. All authors co-wrote the paper.\\

%

\newpage
\setcounter{figure}{0}
\setcounter{table}{0}
\setcounter{equation}{0}

\onecolumngrid

\global\long\def\theequation{S\arabic{equation}}
\global\long\def\thefigure{S\arabic{figure}}
\renewcommand{\thetable}{S\arabic{table}}
\renewcommand{\arraystretch}{0.6}

\normalsize

\vspace{1.0cm}
\begin{center}
{\bf \large Supplementary information for
``A tunable Josephson platform to explore many-body quantum optics in
circuit-QED''}
\end{center}

\subsection{Experimental setup}

\begin{figure}[ht]
\begin{center}
\includegraphics[scale=1.8]{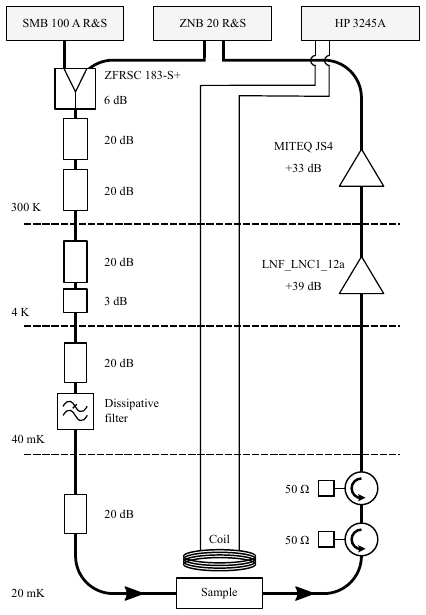}
\caption{{\bf Experimental setup.}}
\label{fig:experimental_setup}
\end{center}
\end{figure}

The full measurement setup is shown in \cref{fig:experimental_setup}. The device
was placed in a dilution refrigerator at a base temperature of \SI{20}{\milli\kelvin}, 
and the transmission measurements were performed using a Vector Network Analyzer (VNA). 
An additional microwave source was used for two-tone measurements, while a global magnetic 
field was applied via an external superconducting coil. 
Both the coil and the sample were held inside a mu-metal
magnetic shield which is coated on the inside with a light absorber made out of
epoxy loaded with silicon and carbon powder. The output line included two
isolators  at \SI{20}{\milli\kelvin}, a HEMT amplifier at \SI{4}{\kelvin} and a
room temperature amplifier. The input line is attenuated at various stages,
including a home-made filter that prevents stray-radiations from reaching the sample. 
We adopted a coaxial geometry with a dissipative dielectric (reference RS-4050 from resin
systems company). The bandwidth of the measurement setup goes from
\SIrange{2.5}{13}{\giga\hertz}. 

\subsection{Chain dispersion relation}

\begin{figure}[ht]
\begin{center}
\includegraphics[scale=0.3]{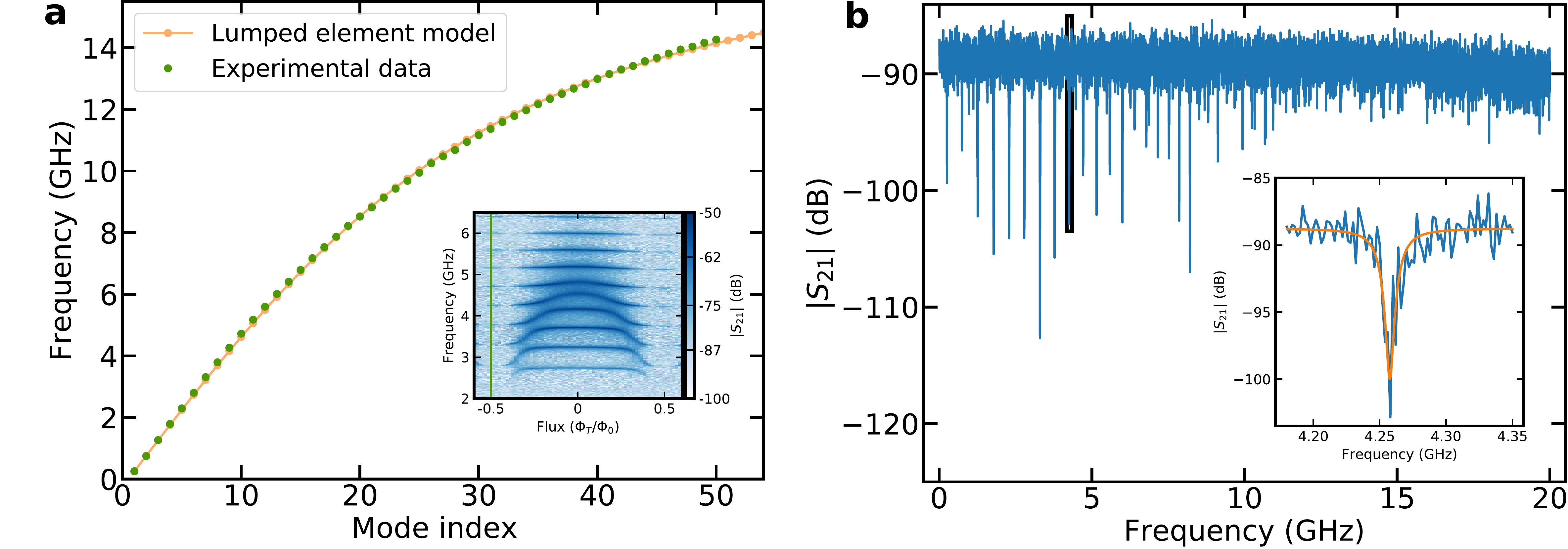}
\caption{{\bf Extraction of the individual chain modes.} \textbf{a}~ Dispersion relation of the chain, reproduced from
Fig.~2\textbf{b} in main text. The inset shows a colourscale plot of the transmission
amplitude as function of $\fluxT$ and probe frequency, with a green line indicating the 
fixed flux value employed to determine the dispersion relation of the uncoupled chain. 
 \textbf{b}~Two tone measurement of the modes of the array. The frequency trace is along the 
 green line in panel {\it a}.
The inset shows a lorentzian fit of one of the dips. The measurement was taken
with $P_{\text{VNA}} = -10$ dBm and $P_{\text{MW}} = 10$ dBm at room temperature. 
The frequency of the VNA was set to $\wVNA = 5.1692$ GHz.} \label{fig:dispersion_relation_measurement}
\end{center}
\end{figure}

In this section we explain how we experimentally obtained the dispersion relation of the chain, and how we used it to determine the model parameters $\Ljmin$, $\Cg$ and $\Cj$ that characterize the chain. (See {\it Methods - Sample fabrication and parameters} in the main text.)
The experimental data in
\cref{fig:dispersion_relation_measurement}\textbf{a} (reproduced from
Fig.~2\textbf{b} in main text) was obtained by first tuning the external magnetic field to 
a point where $\fluxT=\flux_0/2$ so that $\EjQ(\fluxT) = 0$. (Green dashed line in the inset to
\cref{fig:dispersion_relation_measurement}\textbf{a}.) This leads to vanishingly low transmon frequency. As a
result, the transmon does not contribute any degrees of freedom that can
hybridize with the bare modes of the chain. In order to
realize a good fit, one needs to measure the spectrum in a wide frequency band
(\SIrange{0.1}{20}{\giga\hertz}). One is however limited by the bandwidth of the
setup (\SIrange{2.5}{13}{\giga\hertz}). This difficulty can be overcome by performing 
a two-tone measurement\cite{SAnonymous:2012jo, Weissl2014, SWeissl:2015do}, taking advantage of the fact 
that the array is not perfectly linear. As a consequence, when applying a microwave tone 
at a given resonance of the chain, the other resonant frequencies are shifted by the 
cross Kerr effect. With the Vector Network Analyzer (VNA), we proceed by
measuring the transmission of the system at a fixed frequency $\wVNA = \omega_1$ 
where $\omega_1$ matches a given resonance frequency of the circuit. 
Then with a microwave source we apply a second tone at a variable
frequency $\wSMB$. Whenever $\wSMB$ equals any other resonance frequency of the
circuit, $\omega_1$ shifts to $\tilde{\omega}_1$ due to the cross Kerr effect,
so that $\wVNA \neq \tilde{\omega}_1$, which leads to a dip in transmission. The value 
of $\wSMB$ at these dips provides all the resonances of the system. Because we measure 
at a constant frequency $\wVNA$ inside the setup bandwidth, we are not limited by the
frequency range of our measurement setup anymore. A typical two-tone measurement
is shown in \cref{fig:dispersion_relation_measurement}\textbf{b}, where we fit each 
dip separately with a Lorentzian. The center frequencies obtained from these fits are the
experimental points in \cref{fig:dispersion_relation_measurement}\textbf{a}
for the eigenmodes of the chain.

In order to fit the experimental data for the chain modes, we assume that the left end 
of the chain is open when $\fluxT = 0.5\ \flux_0$ ($\EjQ{} = 0$).
We also take the right end of the chain to be grounded. Given that the chain SQUIDS are designed
to have a Josephson energy several thousand times their charging energy ($\Ej{}/\Ec{} = 8400$),
we can model the Josephson junctions in
the chain as linear inductors with inductance $\Lj(\fluxC)=\MagQuanreduced^2/E_J(\fluxC)$, with
$\MagQuanreduced=\hbar/2e$ the reduced flux quantum. 
The theoretical dispersion relation can be obtained
applying Kirchoff's laws to one chain cell of length $\lencell{}$. (See the circuit diagram in Fig. 1\textbf{a} in
the main text.) Denoting the
flux at node $\island$ as $\flux_{\island}$, we obtain
\begin{equation}
\frac{1}{\Lj(\fluxC)}\left(\flux_{\island-1}-\flux_\island\right)+\Cj\left(\ddot{\flux}_{\island-1}-\ddot{\flux}_\island\right) 
- \frac{1}{\Lj(\fluxC)}\left(\flux_\island-\flux_{\island+1}\right)-\Cj\left(\ddot{\flux}_\island-\ddot{\flux}_{\island+1}\right)
-\Cg\ddot{\flux}_\island = 0.
\label{eq:chain_kirchoffs}
\end{equation}
Now if we use as ansatz a plane waves $\flux_\island =A \exp i\left(\omega t -
\kappa \island\lencell{}\right) + B \exp i\left(-\omega t +
\kappa \island\lencell{}\right)$ and solve for $\omega$ we obtain the dispersion
relation for a bare chain
\begin{equation}
\omega\left(\kappa\right) =
\frac{1}{\sqrt{\Lj(\fluxC)\Cj}}\sqrt{\frac{1-\cos{(\kappa\lencell)}}{1-\cos{(\kappa\lencell)}
+ \frac{\Cg}{2\Cj}}}.
\label{eq:dispersion_relation_lumped}
\end{equation}
The boundary conditions at site 0 (vacuum) and at site $\Nj$ (grounded) read
\begin{align}
\left.\frac{\partial \flux_\island}{\partial
(\island\lencell)}\right|_{\island=0} &= 0, \\
\dot{\flux}(\Nj) &= 0,
\end{align}
which restricts the values of $\kappa\lencell$ to
\begin{equation}
\kappa\lencell = \frac{\left(\mode-\frac{1}{2}\right)\pi}{\Nj} \hspace{1cm}
\mode = 1,2,\ldots,\Nj.
\label{eq:kappa_values}
\end{equation}
Since the areas of chain SQUID loops are much smaller than that of the transmon SQUID loop,
at $\fluxT=\flux_0/2$ we can tune the flux $\fluxC$ through each chain SQUID to a multiple of $\flux_0$, so that the chain SQUID
inductance is minimal, i.e. $\Lj(\fluxC)=\Ljmin$, without appreciably changing the transmon flux $\fluxT$ from its value $\flux_0/2$. (See Eq.\,(3) in the {\it Methods} section of the main text.)
Using \cref{eq:dispersion_relation_lumped} with the $\kappa$ values from
\cref{eq:kappa_values}, we fit the experimental data in
\cref{fig:dispersion_relation_measurement}\textbf{a} (orange curve), thus fixing
the minimal Josephson inductance $\Ljmin$ and the capacitance to ground $\Cg$. We obtain
the Josephson self-capacitance $\Cj$ using the empirical formula\cite{Fay2008}
\begin{equation}
\Cj = \SI{45}{\femto\farad\per\micro\meter\squared}\times\text{junction area}.
\label{eq:capacitance_area_relation}
\end{equation} 
The error for $\Cj$ is just the error we obtain for the measurement of the
Josephson junction's area using a Scanning Electron Microscope. The error for $\Ljmin$ and $\Cg$ are the values where the deviation between the experiment and
the fit was below \SI{5}{\percent}.

\subsection{Additional phase shift data}

In this section we present a further selection of relative phase shift data
$\delta\phi_n$  obtained for various ($\fluxT$,$\fluxC$) combination. This is
only a small subset of the full data set, and the agreement between theory and
experiment exhibited here is representative of the full data set. Results
presented here complement Fig. 4{\bf a} of the main text. The parameters used to
obtain the theory curves are the ones of \cref{Stab:sample_parameters}.

\begin{figure}[ht]
\begin{center}
\includegraphics[scale=0.25]{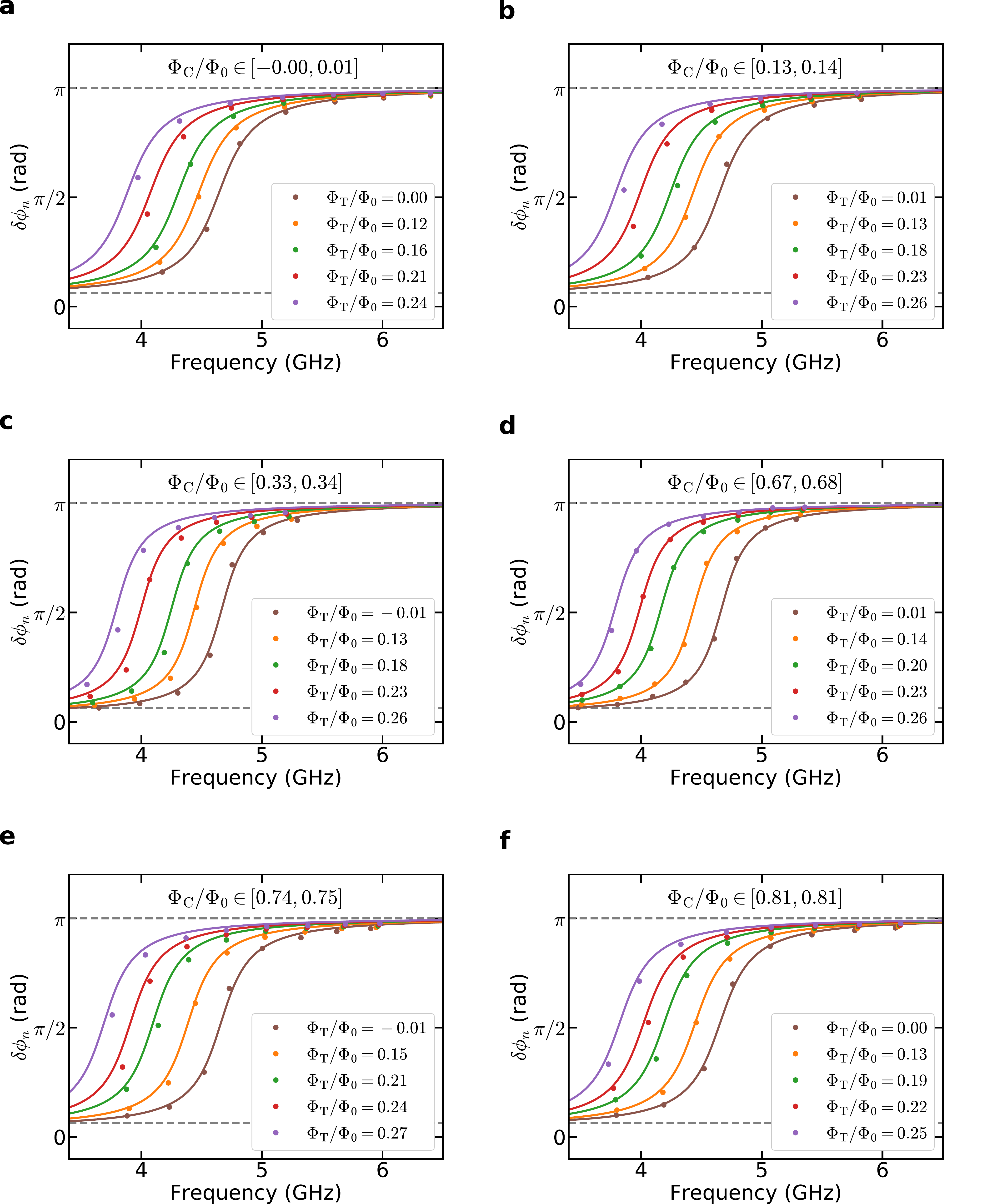}
\caption{{\bf Consistency of the theoretical model for several transmon and
chain fluxes.} The various panels show the relative phase shift $\delta\phi_n$ of the discrete chain modes as
a function of mode frequency $\omega_n$ for different transmon fluxes $\fluxT$ and chain fluxes $\fluxC$. The solid lines are 
fits using Eq. (14) of the main text with the parameters of the circuit kept fixed.}
\label{fig:add_data}
\end{center}
\end{figure}

\subsection{Transmon qubit capacitances estimation}

In order to obtain the capacitances listed in \cref{Stab:sample_parameters}
(reproduced from the main text) we use EM simulation software (Sonnet). This software
solves Maxwell's equations in three dimensions for the specified design of our
device and gives the scattering parameters of the system as a function of frequency. We simulate two parts of the design independently, the interdigital capacitors and the SQUID of the transmon qubit.
%
\begin{table}[htbp]
\footnotesize
\hspace{-1.9cm}
	\parbox{.30\linewidth}{
		\begin{tabular}{l r}
		\hline
		\multicolumn{2}{c}{\textbf{Chain parameters}} \\
		$\Ljmin$ & \SI{0.33 \pm 0.02}{\nano\henry} \\
		$\Cg$ & \SI{0.13 \pm 0.01}{\femto\farad} \\
		$\Cj$ & \SI{259 \pm 14}{\femto\farad} \\
		$\Nj$           & \num{4700}                    \\
		$\asymmetry$ (asymmetry)            & \num{0.25}                    \\
		\hline
		\end{tabular}
	}
\hspace{1.7cm}
	\parbox{.30\linewidth}{
		\begin{tabular}{l r}
		\hline 
		\multicolumn{2}{c}{\textbf{Transmon qubit parameters}} \\
		$\CgQs$   & \SI{33 \pm 1}{\femto\farad} \\
		$\CgQ$  & \SI{48 \pm 2}{\femto\farad} \\
		$\Cc$  & \SI{119 \pm 2}{\femto\farad} \\
		$\Csh$  & \SI{6.9 \pm 0.1}{\femto\farad} \\
		$\CjQ$  & \SI{5.2 \pm 0.3}{\femto\farad} \\
		$E_{J,\text{T,max}}/h$   & \SI{10.2 \pm 0.4}{\giga\hertz} \\
		$\EcQ/h$ & \SI{2.4 \pm 0.1}{\giga\hertz} \\
		\hline
		\end{tabular}
	}
\caption{{\bf Sample parameters.}}
\label{Stab:sample_parameters}
\end{table}
\subsubsection{Interdigital capacitors}
 
Since we are only interested in modelling the capacitors of the
transmon, we remove the chain from the simulation and replace the Josephson
junction of the transmon by a linear inductor, $\LjSonnet$. We place two ports
at both ends of the design, and set the characteristic impedance of the port on
the left to $\Zleft$ and for the port on the right to $\Zright$. 

\begin{figure}[ht]
\begin{center}
\includegraphics[scale=0.3]{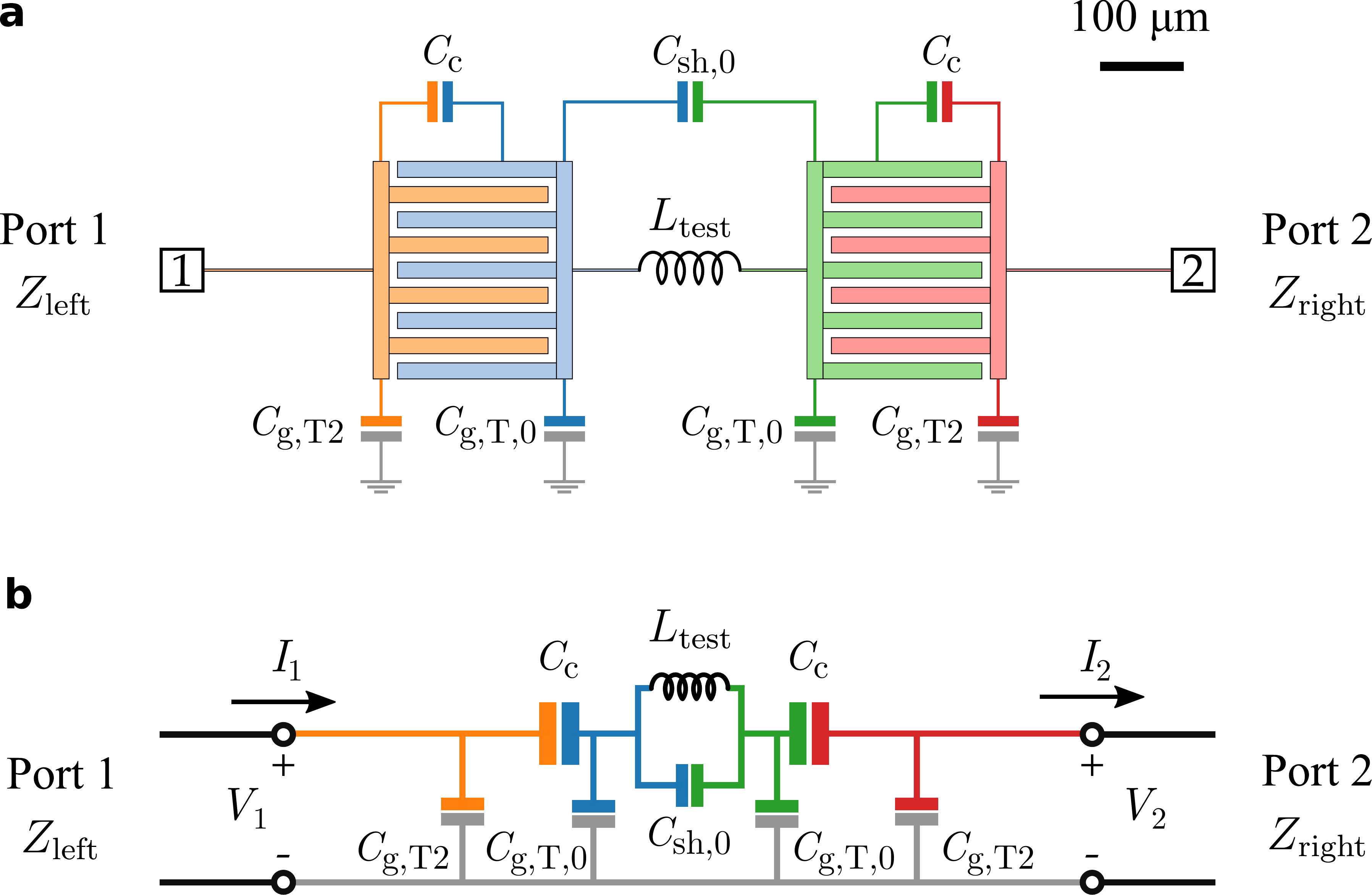}
\caption{{\bf Model for the linear transmon coupling.} \textbf{a}~Real capacitor design.\textbf{b}~Lumped element model used in the EM simulations.}
\label{fig:sonnet_circuit}
\end{center}
\end{figure}

From the EM simulation we obtain the transmission of the system, $S_{21}$, as a
function of frequency. We fit the prediction of the linear model of the qubit to this.
This model consists of the capacitance network shown in
\cref{fig:sonnet_circuit} in red. 
The transmission of this system is given by
\cref{eq:transmission_Sonnet}\cite{Frickley1994} where $A$, $B$, $C$ and $D$ are
the ABCD matrix elements\cite{Pozar2005} for the capacitance network plus the
linear inductor.
\begin{equation}
S_{21} = \frac{2\sqrt{\Zleft\Zright}}{A\Zright + B + C\Zleft\Zright + D\Zleft}.
\label{eq:transmission_Sonnet}
\end{equation}

In theory $\LjSonnet$, $\Zleft$, $\Zright$ do not affect the obtained capacitances and can
be chosen arbitrarily. However, due to the fact that the lumped element model is an idealization, 
we observed a small shift of the capacitances as a function of $\LjSonnet$. (This
shift was not observed as a function of $\Zleft$ or $\Zright$). To minimize the
effect of this shift, we set $\LjSonnet = \SI{22}{\nano\henry}$ which gives a
resonance frequency close to $\Qfreq{}$.

\begin{figure}[ht]
\begin{center}
\includegraphics[scale=0.3]{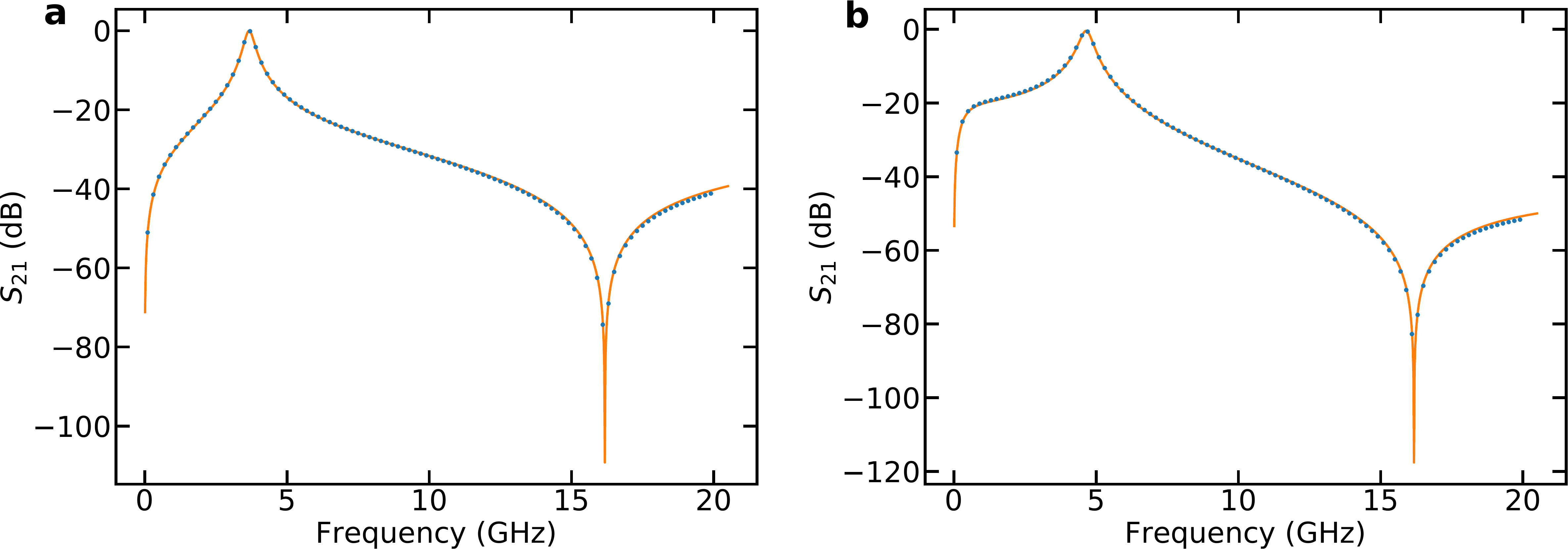}
\caption{{\bf Extraction of the coupling capacitances.} Fit of the obtained 
$S_{21}$ parameter to a linear model with
$\Zright = \SI{50}{\ohm}$ (left panel \textbf{a}) and  $\Zright =
\SI{3000}{\ohm}$ (right panel \textbf{b}).}
\label{fig:sonnet_fit}
\end{center}
\end{figure}

We perform two simulations. In the first one we set $\Zleft = \Zright =
\SI{50}{\ohm}$. Due to the low impedance of the ports, we can neglect $\CgQs$
and we therefore fit only $\Cc$, $\CgQEM$ and $\CshEM$. Then we perform a second
simulation with $\Zleft = \SI{50}{\ohm}$ and $\Zright = \SI{3000}{\ohm}$. Now we
fit only $\CgQs$ keeping the other capacitances constant. In this way we obtain
all the capacitances in \cref{fig:sonnet_circuit} independently. Note that
the self-capacitance of the junction $\CjQ$ cannot be simulated and is therefore
obtained from \cref{eq:capacitance_area_relation}. The errors are obtained as
the maximum range where the difference between simulation and model is smaller
than \SI{10}{\percent}. The two fits are shown in \cref{fig:sonnet_fit}.

\subsubsection{Stray capacitances from the transmon SQUID}

\begin{figure}[htbp]
\centering
\includegraphics[scale=0.3]{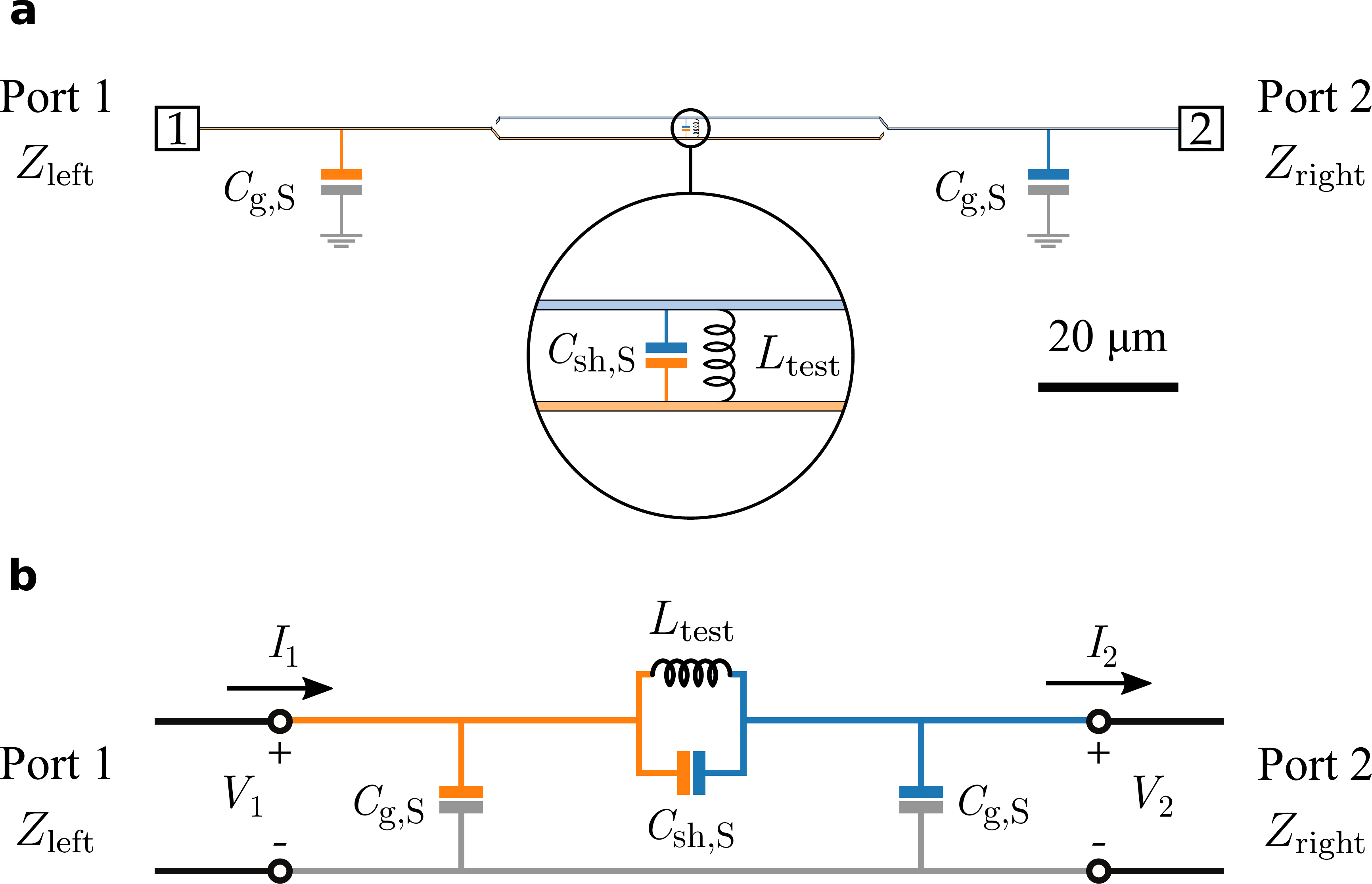}
\caption{{\bf Internal capacitance model for the transmon.} \textbf{a}~Real design of the SQUID of the transmon qubit. In the inset the shunting capacitance $\CshSquid$ and test inductance $\LjSonnet$ are shown. \textbf{b}~Lumped element model used to simulate the capacitances of the system.}
\label{fig:sonnet_SQUID}
\end{figure}

The transmon qubit has a SQUID with a large loop ($\sim \SI{55 x 1.2}{\micro\meter}$). Due to its large size, the capacitances associated to this SQUID are not negligible. In \cref{fig:sonnet_SQUID}\textbf{a} the SQUID design with the different capacitances is given. The lumped element model used for simulating the system is shown in \cref{fig:sonnet_SQUID}\textbf{b}. 

\begin{figure}[htbp]
\centering
\includegraphics[scale=0.3]{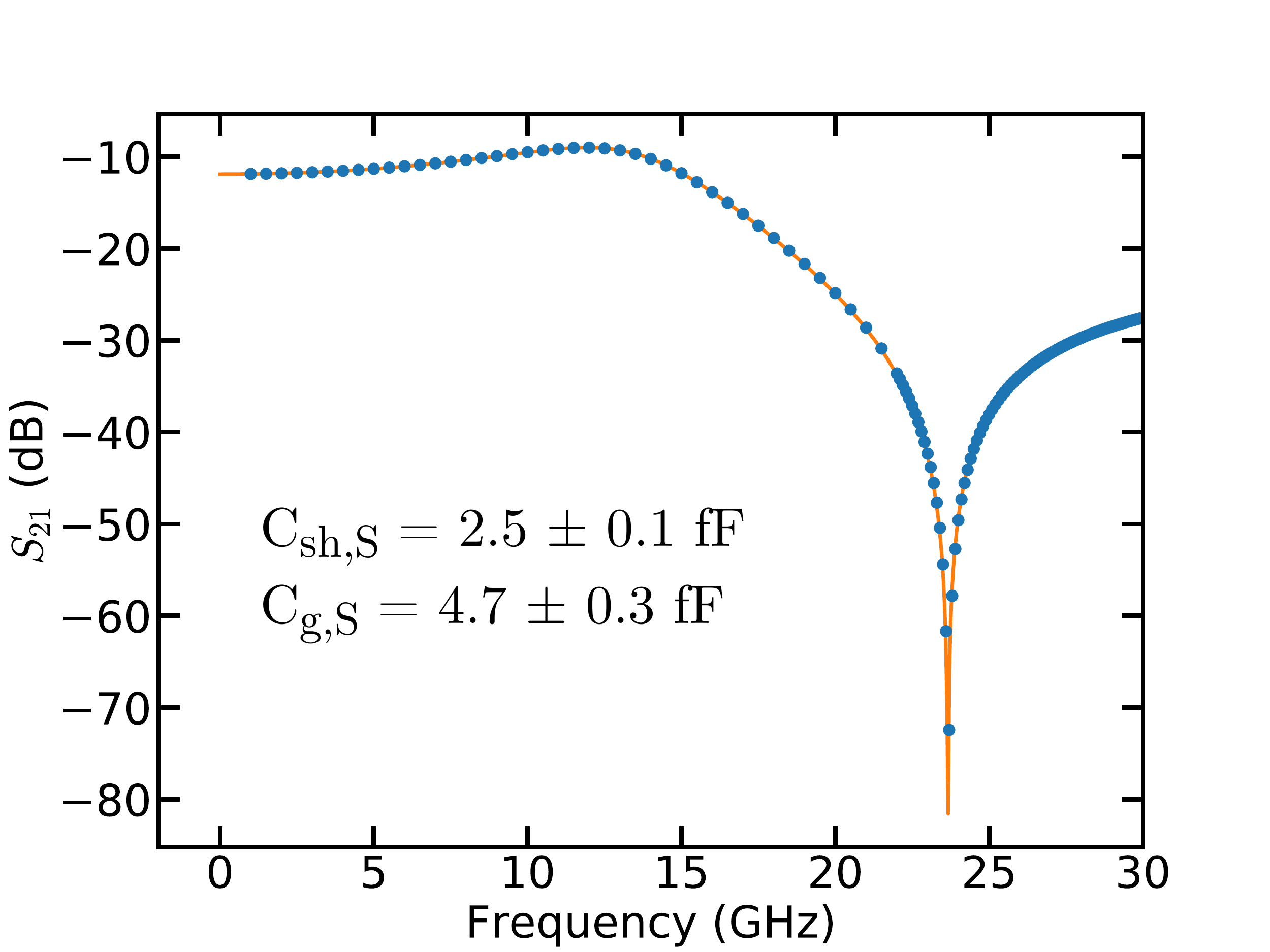}
\caption{{\bf Extraction of the internal capacitances.} SQUID capacitances estimation. Simulated transmission (blue circles) and fit from the lumped element model, orange solid line. The simulation was performed setting $\Zleft = \SI{50}{\ohm}$ and $\Zright = \SI{3000}{\ohm}$.}
\label{fig:sonnet_fit_SQUID}
\end{figure}

We follow the same procedure as before. Given the small number of fitting parameters ($\CshSquid$ and $\CgSquid$) we can perform a single fit with $\Zleft = \SI{50}{\ohm}$ and $\Zright = \SI{3000}{\ohm}$. The result of the fit is shown in \cref{fig:sonnet_fit_SQUID} with the obtained capacitance values.
The SQUID increases both the shunting capacitance and the ground capacitance of the transmon qubit.

\subsection{How does the transmon decay rate depends on the impedance of the environment?}

\begin{figure}[ht]
\begin{center}
\includegraphics[scale=0.6]{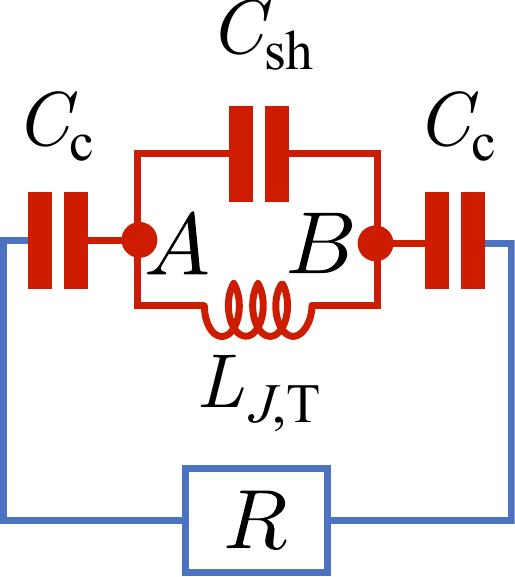}
\caption{{\bf Toy model of the whole circuit.} Simplified lumped element model used to qualitatively understand the link between the transmon decay rate and the impedance of the environment.}
\label{fig:simplified_circuit}
\end{center}
\end{figure}

Before quantitatively modeling the system, in this section we try to gain a qualitative
understanding of how the transmon decay rate $\Gamma_\text{T}$ depends on the 
characteristic impedance of the chain. Since we are only aiming for a qualitative description, 
we treat the transmon SQUID loop as an LC circuit, ignoring its non-linearity. 
We retain the capacitive couplings $\Cc$, that couple the transmon to the chain and the $50$ $\Omega$ transmission line. We drop the ground capacitances $\CgQ$ and $\CgQs$ that shunt the chain  at high frequencies, thus idealizing to the situation where
the chain produces an optimal broadening of the transmon resonance. We consider an infinite chain.
Since we are not interested here in modeling frequency dependent transport through the system, 
but only in the effect
the chain has on the transmon, we replace the complicated frequency dependent impedance of the
chain $Z_\text{chain}(\omega)$ with its constant characteristic
impedance $R=\sqrt{\Lj/\Cg}$. We replace the $50$ $\Omega$ (low impedance) transmission lines
by ground connections. These assumptions produce the simple linear circuit depicted in \cref{fig:simplified_circuit}.

Within this linear model, the resonance frequency $\omega_\text{T}$ and decay rate $\Gamma_\text{T}$
of the transmon are obtained as 
respectively the real and imaginary parts of the relevant pole of the frequency dependent impedance
between A and B in the circuit diagram. This impedance is given by
\begin{equation}
Z_{AB}(\omega)=\frac{i\LjQ\omega(2+i\Cc R\omega)}{(2+iR\Cc \omega)(1-\LjQ \Csh \omega^2)- \LjQ \Cc\omega^2}
\label{eq:ZAB01}
\end{equation}
We have to note here that we have oversimplified the model, which now predicts an overdamped 
regime at small $\Csh$. In the real device, overdamping is prevented by the sizable
capacitances $\CgQ$ and $\CgQs$, which we have dropped. A quick fix, is to use
a value for $\Csh$ that is comparable to $\CgQ$ and $\CgQs$ (several tens of fF), 
rather than its actual value of $4.4$ fF. The behavior of the resonance frequency is easy to understand. 
At small $R$, one effectively has an LC circuit with capacitance $\Csh+\Cc/2$ and resonance frequency
$\omega_\text{T}=1/\sqrt{\LjQ(\Csh+\Cc/2)}$, while at large $R$, one has an isolated SQUID loop with resonance
frequency  $\omega_\text{T}=1/\sqrt{\LjQ \Csh}$.

\begin{figure}[ht]
\begin{center}
\includegraphics[scale=0.4]{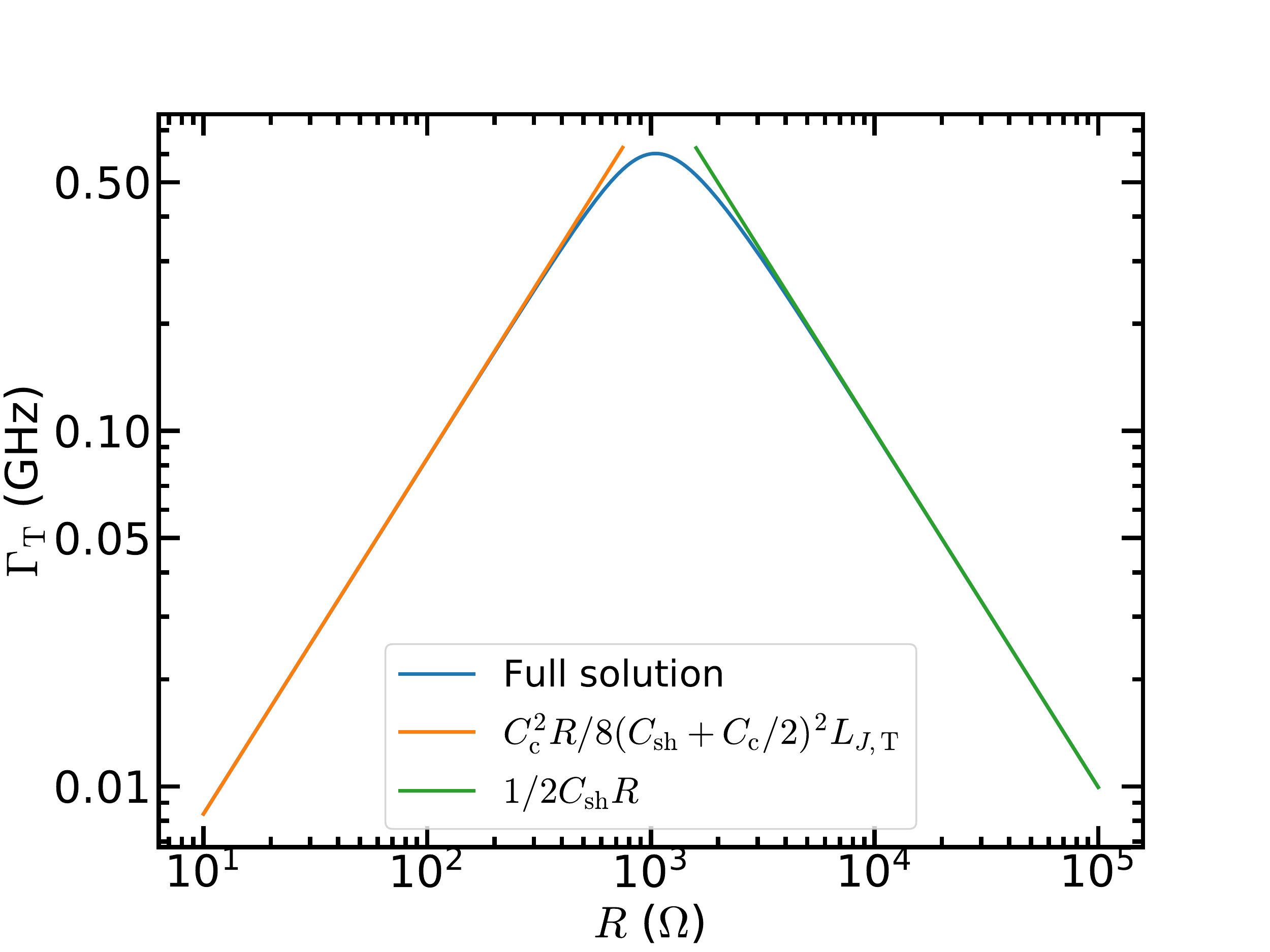}
\caption{{\bf Bath engineered dissipation from the toy model.} Evolution of the transmon decay rate $\Gamma_\text{T}$ versus the value of the resistance $R$, as described in the simplified circuit of \cref{fig:simplified_circuit}.}
\label{fig:transmon_decay_rate}
\end{center}
\end{figure}
   
In \cref{fig:transmon_decay_rate} we present the behavior of $\Gamma_\text{T}$ vs. $R$ for $\Cc=119$ fF (its actual value) and $\Csh=80$ fF, 
chosen to give rough quantitative agreement with the experimental results we present in the Fig. 4
of the main text, although the qualitative behavior does not change if we change $\Csh$ moderately.
 At small $R$, the behavior of the
decay rate $\Gamma_\text{T}$ is 
\begin{equation}
\Gamma_\text{T}=\frac{\Cc^2 R}{8 (\Csh+\Cc/2)^2\LjQ}
\end{equation}
(proportional to $R$) while at large $R$, it is given by
\begin{equation}
\Gamma_\text{T}=\frac{1}{2\Csh R}
\end{equation}
(proportional to $1/R$). A good estimate of the value $R_*$ that maximizes $\Gamma_\text{T}$ is obtained
by equating the small and large $R$ asymptotic expressions for $\Gamma_\text{T}$. This yields
\begin{equation}
R_*=\left(1+\frac{2 \Csh}{\Cc}\right)Z_\text{T,simp}
\end{equation}
where $Z_\text{T,simp}=\sqrt{\LjQ/\Csh}$ is the characteristic impedance of the transmon, in the simplified circuit of \cref{fig:simplified_circuit}. Thus the largest
coupling (as measured by $\Gamma_\text{T}$) is obtained when the characteristic impedances of the
transmon and chain match up to factors of order one, a result that is familiar in microwave engineering. 
At this optimal chain impedance, the decay rate $\Gamma_\text{T}$ is proportional to $\omega_\text{T}$
with a proportionality constant that is a function of $\Cc/\Csh$. This constant can reach values of order one, implying ultra-strong coupling is attainable. In our actual device, we find $\Gamma_\text{T}$ to
be a decreasing function of $\Lj$, in the $\Lj$ window to which we have access, suggesting that the
lowest chain impedance that we can reach, is larger than the optimal value $R_*$. 
When we compare the characteristic impedances  $Z_\text{T}=\hbar/(2e)^2\sqrt{2\EcQ/\EjQ} \simeq
\SI{760}{\ohm}$ (transmon) and $Z_\text{C}=\sqrt{\Lj/\Cg}\simeq \SI{1590}{\ohm}$ (chain) of
the actual device, we see that indeed $Z_\text{C}>Z_\text{T}$. Note that here we took a realistic estimate for the transmon impedance, that includes the effect of the capacitances $\CgQ$ and $\CgQs$ which were dropped in our qualitative model. Had we naively taken 
$Z_\text{T}=\sqrt{\LjQ/\Csh}$ we would have obtained $Z_\text{T}=1990$ $\Omega$, which though still close to the chain impedance, might have lead us to expect to  observe a maximal value for $\Gamma_T$ as we sweep the $\Lj$ window to which we have access.

\subsection{Dealing with the transmon nonlinearity}
\label{sec:scha}
The results in Fig. 3 in the main text confirm that the transmon qubit is a non-linear quantum
circuit element that is strongly coupled to the chain. Here we review a standard way to deal
with this anharmonicity.\cite{Koch2007} The method is known as the self-consistent harmonic approximation (SCHA) because the
anharmonic term is replaced by a harmonic one whose magnitude is determined self-consistently, via the variational principle. 
We also determine the regime of validity
of the approximations we introduce.  Let us consider the complete Hamiltonian of the
device, neglecting only the weak non-linearity in the chain elements:
\begin{equation}
H = \frac{\EcQ{}}{2}\nop_\text{T}^2-\EjQ{}\cos(\phaseop_\text{T})+\frac{(2e)^2}{2}\sum_{\island\iisland=1}^{\Nj}\nop_\island\left[\Cm^{-1}\right]_{\island,\iisland}\nop_{\iisland}
+\frac{\Ej}{2}\sum_{\island=1}^{\Nj}\left(\phaseop_{\island+1}-\phaseop_{\island}\right)^2+\nop_\text{T}\sum_{\island=1}^\Nj\nu_\island\nop_\island.
\label{Shamiltonian}
\end{equation}
Here we found it convenient to define an operator $\phaseop_{\Nj+1}\equiv 0$ which is not an extra degree of freedom, but simply the zero operator.
To shorten notation we don't indicate the $\fluxT$ dependence of $\EjQ$ or the $\fluxC$ dependence
of $E_J$ explicitly here.
Were it not for the term $-\EjQ{}\cos(\phaseop_\text{T})$, the quantum system described
by the Hamiltonian in \cref{Shamiltonian} (Eq.~8 in the main text) would have
been equivalent to a set of coupled harmonic oscillators, and therefore
straightforward to solve.
The term $-\EjQ{}\cos(\phaseop_\text{T})$, not being quadratic in $\phaseop_\text{T}$,
produces an interacting many-body problem. 
The strategy will be to replace the transmon terms in $H$ with more tractable,
yet accurate counterparts. For this purpose our starting point is to consider
the Hamiltonian $H$ in the limit where the inductances  $\Lj$ between chain
nodes go to infinity ($\Ej\to0$), so that the charge on each chain island is conserved, and
we can treat $\nop_\island$, $\island\in\{1;\,\ldots;\, N\}$ as ordinary
numbers.  Since the transmon then does not couple to any dynamical degrees of
freedom, we refer it as isolated. The conserved chain charges contribute to the
offset charge for $\nop_\text{T}$. We will abuse notation slightly and still
denote the transmon's charge degree of freedom, which now incorporates this
additional offset, by $\nop_\text{T}$. The isolated transmon Hamiltonian, in
which reference to the conserved charges $\nop_\alpha$, 
$\alpha\in\{1;\,\ldots;\, N\}$ has been eliminated, reads
\begin{equation}
H_\text{T}=\frac{\EcQ}{2} \nop_\text{T}^2+ \EjQ [1-\cos(\phaseop_\text{T})].\label{eq:isolated_transmon_hamiltonain}
\end{equation}
Due to charge being quantized in units of $2e$, the state space of $H_\text{T}$ is restricted to states $\left|\psi\right>$ for which
\begin{equation}
e^{i 2\pi  \nop_\text{T}}\left|\psi\right>=e^{-i 2\pi n_\text{T}}\left|\psi\right>,\label{bc}
\end{equation}
where the offset charge $n_\text{T}$ (an ordinary number) contains contributions from
the total transmon charge, the charge on each chain island, and from gate
charges.  We denote the eigenbasis of $\nop_\text{T}$ by $\left|\nu\right>$,
i.e.
\begin{equation}
\nop_\text{T}\left|\nu\right>=\nu\left|\nu\right>.
\end{equation}
Owing to (\ref{bc}), $\nu$ is quantized such that
\begin{equation}
\nu+n_\text{T}\in \mathbb{Z}.
\end{equation}
The matrix elements of $H_\text{T}$ in the $\left|\nu\right>$ basis read
\begin{equation}
\left<\nu\right| H_\text{T}\left|\nu'\right>=\frac{\EcQ \nu^2}{2}\delta_{\nu,\nu'}+\frac{\EjQ }{2}\left(\delta_{\nu,\nu'+1}+\delta_{\nu,\nu'-1}\right).
\end{equation}  
\begin{figure}[ht]
\begin{center}
\includegraphics[width=.32 \textwidth]{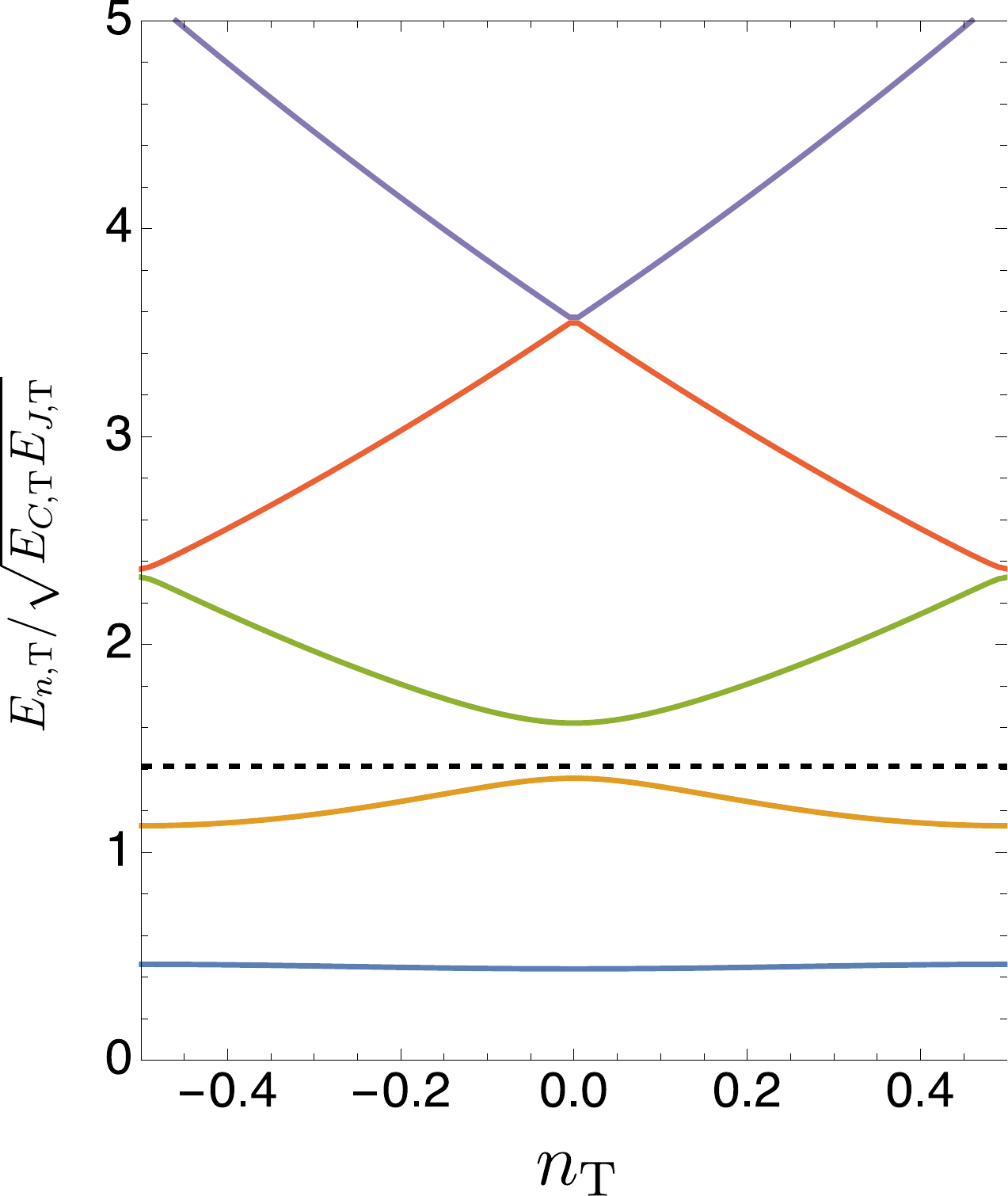} 
\includegraphics[width=.32 \textwidth]{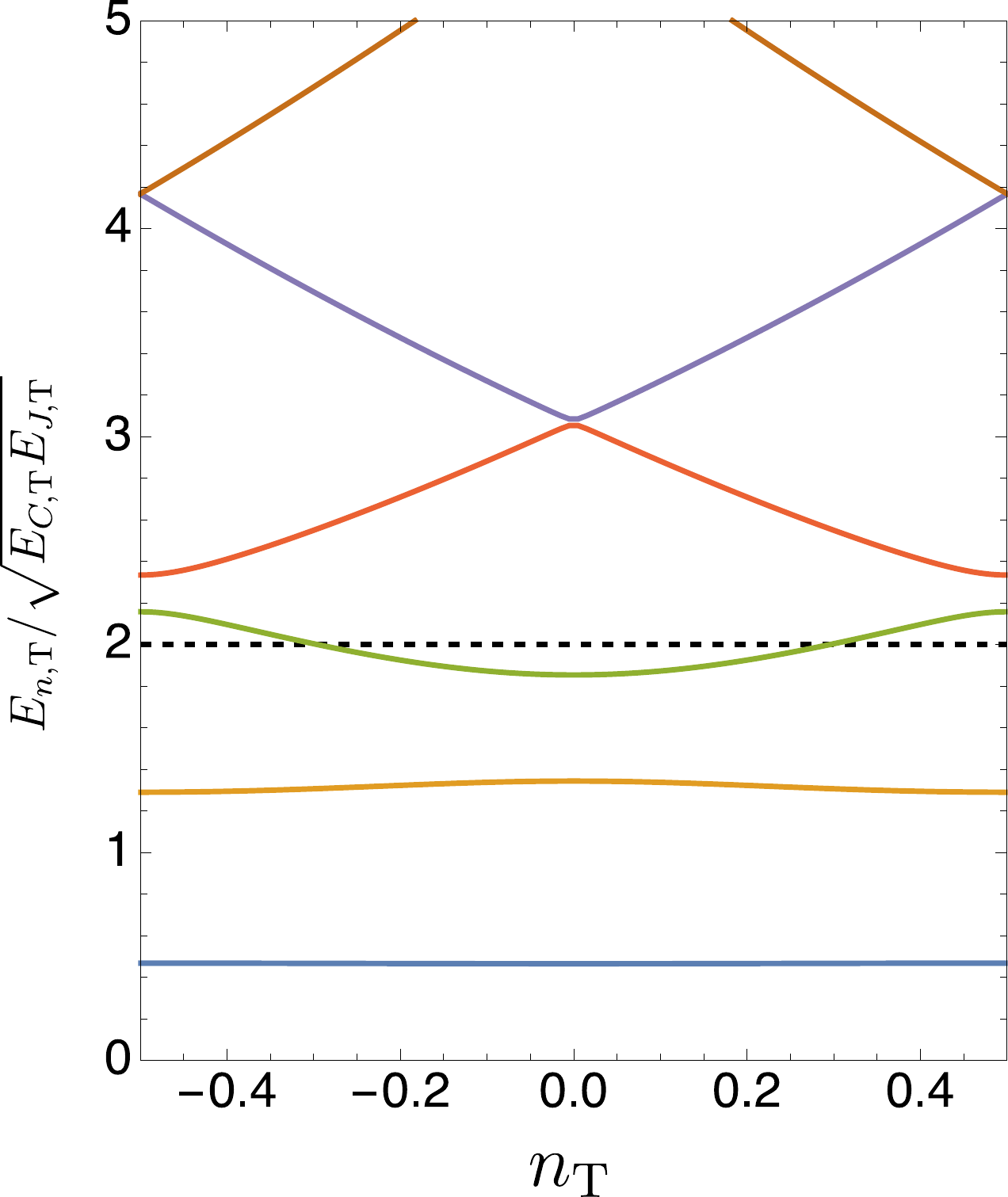} 
\includegraphics[width=.32 \textwidth]{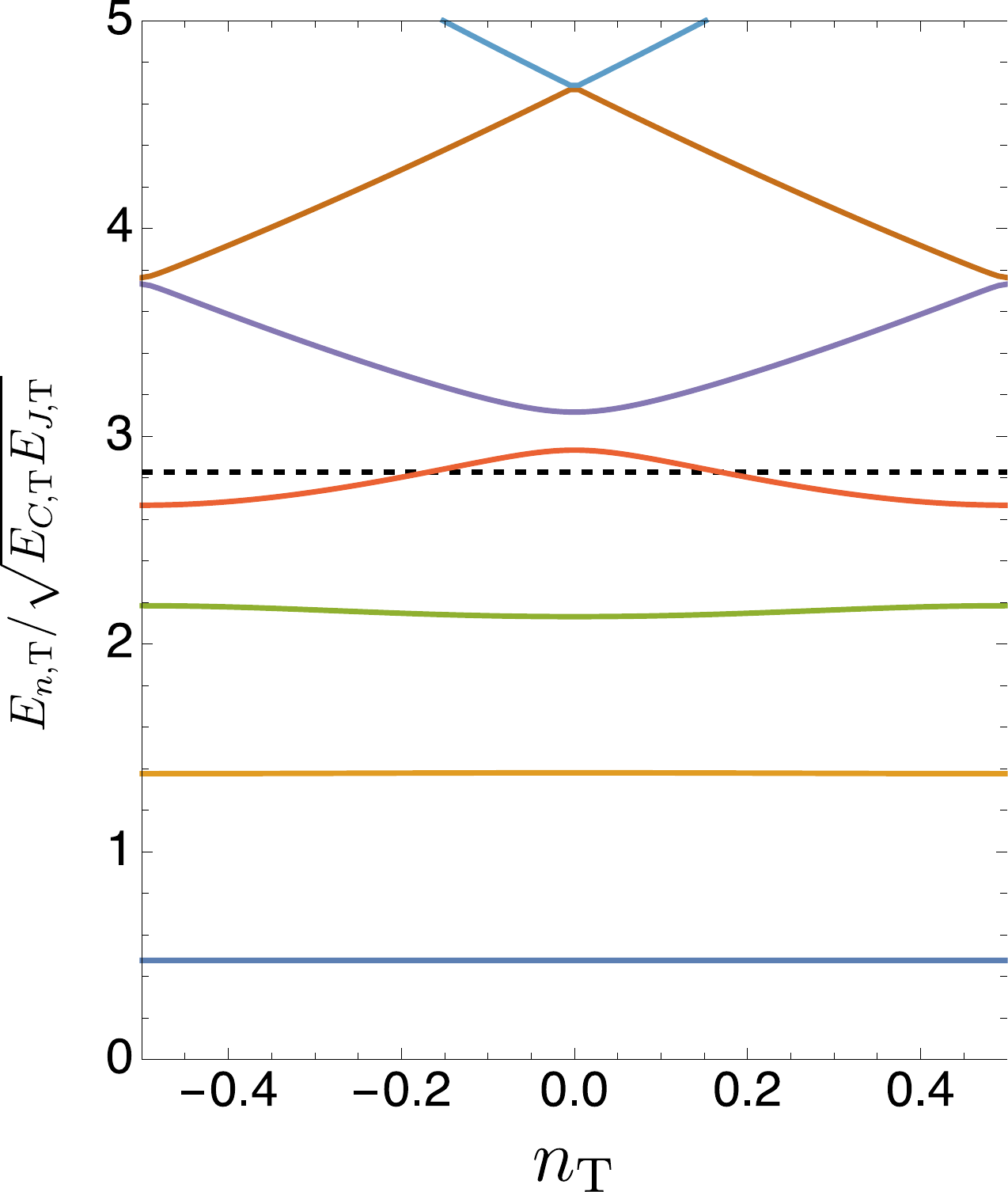} 
\end{center}
\caption{{\bf Isolated transmon spectrum as a function on the offset charge}
$\mathbf{n_\text{T}}$. Left panel: $\EjQ/\EcQ=1/2$. Middle panel: $\EjQ/\EcQ =1$.  Right
panel: $\EjQ/\EcQ=2$. In each case a dashed line indicates the energy $2\EjQ$,
maximum potential energy.  \label{f1}} 
\end{figure}
The problem is equivalent to that of a charge $e$ particle of mass $E_{C, {\rm
T}}^{-1}$ confined to a ring of circumference $2\pi$ that is threaded by a flux
of $n_\mathrm{T}$ times the flux quantum $\MagQuan = h/2e$. The phase observable
$\phaseop_\text{T}$ plays the role of the position coordinate, and the particle
has an electrostatic potential energy $\EjQ(1-\cos \phaseop_\text{T})$. This
system is easily solved numerically. In Figure \ref{f1} the low energy spectrum
is plotted as a function of the offset charge $n_\text{T}$, for three $\EcQ/\EjQ
$ ratios.
States with energies sufficiently less than the height $2 \EjQ$ of the cosine
well are insensitive to the offset charge $n_\text{T}$. This is easy to
understand in the equivalent picture of the particle confined to a ring:
Sensitivity to the flux inside the ring requires the interference of paths with
different winding numbers around the ring. However, at energies below $2 \EjQ$,
paths with non-zero winding number are exponentially suppressed by the tunneling
amplitude to go through the cosine barrier. 
States with energy $\gtrsim 2\EjQ$ on the other hand are sensitive to the offset
charge $n_\text{T}$. As $n_\text{T}$ varies form $0$ to $1/2$ (half a Cooper
pair), each of these energies sweep through an interval (or band) of width
comparable to the spacing between levels. In the equivalent picture of a
particle on a ring, this is a manifestation of the Aharonov Bohm effect. In a
real experiment, the offset charge $n_\text{T}$ is subject to environmental
noise.  Performing spectroscopy on the levels sensitive to $n_\text{T}$ will
therefore produce a noisy signal in which the extracted level energy ``jumps
around'' inside the band through which the energy sweeps as $n_\text{T}$ is
varied.

When we reduce $\Lj$ from infinity to its actual value, thus coupling the
transmon to dynamical degrees of freedom in the chain, a numerically exact
solution is no longer possible, given the large size of the Hilbert space. An
obvious approximation scheme for states with energies below $2\EjQ$ is the
following. For these states, the phase observable $\phaseop_\text{T}$ is
unlikely to make excursions over the top of the cosine barrier (phase slips) at
$\varphi_\text{T}=\pm \pi$. For such states it should therefore be permissible
to replace the ring to which the particle is confined with the whole real line,
and the cosine potential with a parabola.  Note that this approximation ignores
the restriction (\ref{bc}), responsible for charge quantization.  For states
with energies $\ll \EjQ$, the phase is confined very close to the minimum at
$\varphi=0$ of the cosine potential, and the replacement
$\EjQ(1-\cos\phaseop_0)\to \EjQ\phaseop_0^2/2$ is legitimate. This leads to a
harmonic spectrum $\omega_n=\sqrt{\EcQ \EjQ}(n+1/2)$.
However, the quadratic approximation can be improved to have a larger regime of
validity, in the following way. We approximate the eigenstates of the {\bf
isolated} transmon as those of the parent Hamiltonian
\begin{equation}
H_P=\frac{\EcQ }{2} \nop_\text{T}^2 + \frac{E_S}{2}\phaseop_\text{T}^2,
\end{equation}
where the parameter $E_S$ is optimized according to some criterium in order to
give the best possible agreement with the exact solution. Here we use the
criterion that the energy $E_S$ should be chosen to minimize $\left<H_\text{T}\right>$
where the expectation value is taken with respect to the ground state of $H_P$.
For given $E_S$, the eigenstates and energies of $H_P$ are
\begin{equation}
\left|n\right>=\frac{(B^\dagger)^n}{\sqrt{n!}}\left|0\right>,~~~E_n^{(0)}=\Qfreq(n+1/2)
,~~~n=0,\,1,\,2,\,\ldots,~~~\Qfreq=\sqrt{\EcQ  E_S}, \label{zeroorder}
\end{equation}
where
\begin{equation}
B\left|0\right>=0,~~~B=\lambda \nop_\text{T}+i \phaseop_\text{T}/2\lambda
,~~~\lambda=\frac{1}{\sqrt{2}}\left(\frac{\EcQ }{E_S}\right)^{1/4}.\label{eq:Bdef}
\end{equation}
Expressed in terms of the bosonic opertors $B$ and $B^\dagger$, and manipulated
into normal ordered form, the full transmon Hamiltonian reads
\begin{align}
H_\text{T}=&\Qfreq\left[B^\dagger B+\frac{1}{2}\right]+\EjQ\left\{1-\frac{e^{-\lambda^2/2}}{2}\left[e^{\lambda B^\dagger}e^{-\lambda B}+e^{-\lambda B^\dagger}e^{\lambda B}\right]\right\}\nonumber\\
&+\frac{E_S}{2}\lambda^2\left[(B^\dagger)^2+B^2-2(B^\dagger B+1)\right].
\end{align}
The expectation value $\left<0\right|H_\text{T}\left|0\right>$ evaluates to
\begin{equation}
\left<0\right|H_\text{T}\left|0\right>=\frac{\Qfreq}{4}+\EjQ\left(1-e^{-\lambda^2/2}\right)=\frac{\sqrt{\EcQ E_S}}{4}+\EjQ\left[1-e^{-\sqrt{\EcQ /E_S}/4}\right].
\end{equation}
The minimal value for $\left<0\right|H_\text{T}\left|0\right>$ is produced by $E_S$ satisfying the equation
\begin{equation}
E_S=\EjQ e^{-\sqrt{\EcQ /E_S}/4},\label{scon}
\end{equation}
which can also be written as an equation 
\begin{equation}
\Qfreq=\sqrt{\EjQ \EcQ} e^{-\EcQ /8\Qfreq},\label{scon2}
\end{equation}
determining the transmon frequency $\Qfreq$.
We note that in principle, the approximation can be further improved by treating
$H_P$ with the optimized value (\ref{scon}) for $E_S$ as the zero'th order
approximation and treating $H_\text{T}-H_P$ as a small perturbation. In general,
the leading corrections in such a perturbation expansion are of first order in
$\lambda^2$. However, for the ground and first excited states, it is one order
higher, i.e. $\lambda^4$. Thus, the approximation $H_\text{T}\simeq H_P$ is
particularly accurate for the ground and first excited states of the transmon,
which given the probe power and plasma frequency of the chain, are the ones we are interested in, in the experiment. 
\begin{figure}[ht]
\begin{center}
\includegraphics[width=.60 \textwidth]{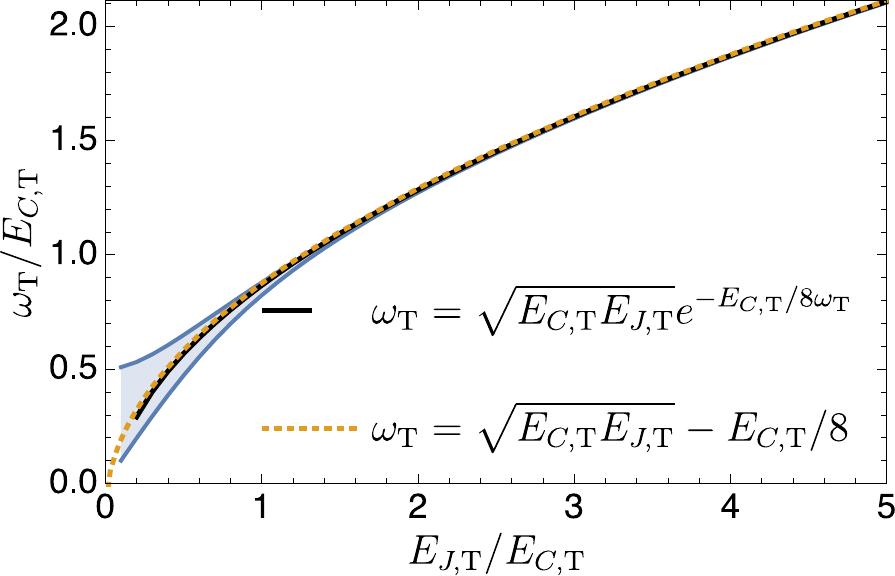}  
\end{center}
\caption{{\bf From the transmon to the Cooper pair box.} Excitation energy $\Qfreq$ from the ground state to the first excited
state of the isolated transmon. The shaded area respresents the range of values
obtained for the exact result when the offset charge $n_\text{T}$ is swept through 
$[0,1/2]$. The solid black line represents the fully self-consistent result (\ref{scon2}) while
the dashed line represents the simple approximation (\ref{approx}) . \label{f3}} 
\end{figure}
The assumed smallness of $\lambda$ allows us to solve approximately the
self-consistency equation (\ref{scon}) to get
\begin{equation}
E_S=\EjQ-\frac{1}{4} \sqrt{\EcQ  \EjQ}+\mathcal O \left(\EcQ \right),\label{appes}
\end{equation}
which gives an excitation energy 
\begin{equation}
\Qfreq=\sqrt{\EcQ  \EjQ}-\EcQ /8+\mathcal O \left(\frac{\EcQ ^2}{\sqrt{\EcQ  \EjQ}}\right).\label{approx}
\end{equation} 
In Figure \ref{f3} we compare the results (\ref{scon2}) and (\ref{approx}) to
the exact excitation energy of the isolated transmon, and find that in the
regime where sensitivity to the offset charge $n_\text{T}$ is weak, i.e.
$\EjQ/\EcQ \gtrsim 1$, the approximation (\ref{approx}) is indistinguishable
from the more sophisticated (\ref{scon2}).
Now we turn to the case $\Ej>0$ where the transmon is coupled to the dynamical
degrees of freedom in the chain.  We take the same approach as before.  In
principle, it is possible to generalize the previous calculation in the
following way.  When $\Ej>0$, we may choose $E_S$ so that it minimizes
$\left<0\right|H\left|0\right>$, where $H$ is the full Hamiltonian in
 (Eq.~4 in the main text)
\begin{equation}
H = \frac{(2e)^2}{2}\vec{\nop}^T\Cm^{-1}\vec{\nop} - \frac{1}{2}\vec{\phaseop}^T\Jm\vec{\phaseop} 
- \EjQ\cos\left(\phaseop_R - \phaseop_L\right),
\end{equation}
and $\left|0\right>$ is the ground state
of the parent Hamiltonian that is obtained from $H$ by making the replacement
$\EjQ(1-\cos\phaseop_\text{T})\to E_S \phaseop_\text{T}^2/2$. 
We have implemented this approach, but find that it yields an insignificant
improvement upon the simpler approach of simply taking $E_S=\EjQ-\sqrt{\EjQ\EcQ
}/4$, at least in the one photon sector, and for the parameters of the current
device.

\subsection{Analytical formula for the Scattering Phase Shift}

In the section {\it Methods -  Analytical formula for the Scattering Phase Shift} we presented
a formula (Eq.\,11) for the phase shift $\phi(\omega,\fluxT,\fluxC)$ of a mode with frequency $\omega$ of the full system at transmon flux
$\fluxT$ and chain flux $\fluxC$:
\begin{equation}
\tan{\phi(\omega,\fluxT,\fluxC)} =
\frac{\Cg-2\Ceff(\fluxT,\omega)}{\sqrt{\Cg(\Cg+4\Cj)}}\frac{1}{\sqrt{\left(\frac{\wplasma(\fluxC)}{\omega}\right)^2-1}}.
\label{Seq:phi_analytic}
\end{equation}
Here we give the derivation. 
We work in the
thermodynamic limit where $\Nj \rightarrow \infty$. This means that both
matrices $\Cm$ and $\Jm$
(\cref{eq:capacitance_matrix_semi_infinite,eq:J_matrix_semi_infinite}) become
semi-infinite. Furthermore, the approximation discussed in the previous section
eliminates the anharmonic term in $H$, and introduces four new non-zero matrix
elements in $\Jm$, namely
\begin{align}
\Jm_{L,L} &= \Jm_{R,R} = \Eself\left(\fluxT\right), \\
\Jm_{R,L} &= \Jm_{L,R} = -\Eself\left(\fluxT\right). 
\end{align}
Making the flux-dependence of $E_S$ explicit, we have from \cref{appes}
\begin{equation}
\Eself\left(\fluxT\right) = \EjQ\left(\fluxT\right) -
\sqrt{\EjQ\left(\fluxT\right)\EcQ}/4,
\label{eq:E_SCHA}
\end{equation}
with $\EjQ\left(\fluxT\right) = E_{J,T,\text{max}}\left|\cos{(\pi\fluxT/\MagQuan)}\right|$.
Explicitly the matrices $\Cm$ and $\Jm$ then read (see Eqs. (6) and (7) in the main text)
\begin{equation}
\Cm = 
	\begin{pmatrix}
		C_{0}  &  -\CshQ     &0         &0         &0          &0            &\cdots\\
		   -\CshQ &   C_{0}  &-\Cc      &0         &0          &0            &\cdots\\
		      0  &-\Cc       &C_{1}&-\Cj      &0          &0            &\cdots\\
			0      &0          &-\Cj       &\Csum&-\Cj        &0            &\cdots\\
			\vdots &\vdots     &\vdots    &\ddots    &\ddots     &\ddots       &\cdots
	\end{pmatrix}
	\label{eq:capacitance_matrix_semi_infinite}
\end{equation}
\begin{equation}
\Jm = \frac{\varphi_0^2}{\Lj\left(\fluxC\right)}
	\begin{pmatrix}
		0  &  0   &0         &0         &0          &0            &\cdots\\
		   0 &   0  &0      &0         &0          &0            &\cdots\\
		      0  &0       &1&-1&0          &0            &\cdots\\
			0      &0          &-1       &2&-1      &0            &\cdots\\
			\vdots &\vdots     &\vdots    &\ddots    &\ddots     &\ddots       &\cdots
	\end{pmatrix}
	+
	\Eself(\fluxT)
	\begin{pmatrix}
		1  &  -1   &0         &0         &0          &0            &\cdots\\
		   -1 &   1  &0      &0         &0          &0            &\cdots\\
		      0  &0       &0& 0&0          &0            &\cdots\\
			0      &0          &0       & 0& 0      &0            &\cdots\\
			\vdots &\vdots     &\vdots    &\vdots    &\vdots     &\vdots       &\ddots
	\end{pmatrix}
	\label{eq:J_matrix_semi_infinite}
\end{equation}
We now have a fully linear system. The phase shifts in \cref{Seq:phi_analytic} are
obtained by solving the classical equations of motion. We start by defining a
vector with the superconducting phases in each island $\vec{\pi}^T =
(\varphi_L,\varphi_R,\varphi_1,\varphi_2,\cdots,\varphi_{\Nj})$. The equations
of motion for the mode at frequency $\omega$ are given by
\begin{equation}
\Jm\left(\fluxT,\fluxC\right)\vec{\pi}_{\omega} =
(\hbar\omega)^2\frac{\Cm}{(2e)^2}\vec{\pi}_{\omega}.
\label{eq:euler_lagrange}
\end{equation}
The solution to the part of \cref{eq:euler_lagrange} involving degrees of freedom in the chain can be taken as 
\begin{equation}
\varphi_{\island} = N(\omega)\cos\left[\island\kappa(\omega,\fluxC)\lencell{} - \phi(\omega,\fluxT,\fluxC) \right] \quad \text{with} \quad \island = 1,2,3,4 \ldots
\label{eq:cosine_mode}
\end{equation}
Here $\lencell{}$ is the length of the unit cell of the array, $N(\omega)$ a frequency dependent amplitude  and
$\kappa(\omega,\fluxC)$ is the wave number of a wave that propagates in the
chain with angular frequency $\omega$. It can be obtained from the dispersion
relation in \cref{eq:dispersion_relation_lumped},
\begin{equation}
\kappa(\omega,\fluxC) =
\frac{2}{\lencell{}}
\mathrm{arccot}\sqrt{\left(\frac{4\Cj}{\Cg}+1\right)\left[\left(\frac{\wplasma(\fluxC)}{\omega}\right)^2
- 1\right]},
\label{eq:wave_vector}
\end{equation}
where $\wplasma(\fluxC) = 1/\sqrt{\Lj(\fluxC)\left(\Cj+\Cg/4\right)}$ is the plasma frequency of
the chain. The phase shift $\phi(\omega,\fluxT,\fluxC)$ in \cref{eq:cosine_mode}
is determined by the components of \cref{eq:euler_lagrange} involving the
transmon islands. They read
\begin{align}
\Eself\left(\fluxT\right)\left(\varphi_L-\varphi_R\right) &=
\frac{(\hbar\omega)^2}{\left(2e\right)^2}\left(C_0\varphi_L-\CshQ\varphi_R\right)
\label{eq:phi_R}, \\ 
\Eself\left(\fluxT\right)\left(\varphi_R-\varphi_L\right) &=
\frac{(\hbar\omega)^2}{\left(2e\right)^2}\left(-\CshQ\varphi_L-C_0\varphi_R-\Cc\varphi_1\right)
\label{eq:phi_L}, \\ 
\Eself\left(\fluxT\right)\left(\varphi_1-\varphi_2\right) &=
\frac{(\hbar\omega)^2}{\left(2e\right)^2}\left(-\CshQ\varphi_R-C_1\varphi_1-\Cj\varphi_2\right)
\label{eq:phi_1}.
\end{align}
Using \cref{eq:phi_R} and \cref{eq:phi_L} we eliminate $\varphi_L$ and solve for
$\varphi_R$ in terms of $\varphi_1$ to obtain
\begin{equation}
\varphi_R = \frac{\Cc\left[\frac{(\hbar\omega)^2}{(2e)^2} C_0 -
\Eself(\fluxT)\right]}{(C_0 -\CshQ)
\left[(C_0+\CshQ)\frac{(\hbar\omega)^2}{(2e)^2}C_0 - 2\Eself(\fluxT)\right]}\varphi_1.
\end{equation}
Substituting this into \cref{eq:phi_1} and using \cref{eq:wave_vector} for $\kappa(\omega,\fluxC)$ we obtain
\begin{equation}
\frac{\Cg}{2\left[1-\cos \kappa(\omega,\fluxC)\lencell{} \right]}(\varphi_1-\varphi_2) = \Ceff(\fluxT,\omega)\varphi_1
\label{eq:phi_1_phi_2},
\end{equation}
with
\begin{equation}
\Ceff(\fluxT,\omega) = C_1 - \Cj -
\frac{\Cc^2\left[\frac{(\hbar\omega)^2}{(2e)^2} C_0 -
\Eself(\fluxT)\right]}{(C_0
-\CshQ)\left[(C_0+\CshQ)\frac{(\hbar\omega)^2}{(2e)^2}C_0 -
2\Eself(\fluxT)\right]}.
\end{equation}
Using the mode definition in \cref{eq:cosine_mode} for $\varphi_1$ and $\varphi_2$ in \cref{eq:phi_1_phi_2} leads to
\begin{equation}
\tan{\phi(\omega,\fluxT,\fluxC)} =
\left[1-\frac{2\Ceff(\fluxT,\omega)}{\Cg}\right]\tan{\frac{\kappa(\omega,\fluxC)\lencell{}}{2}}.
\end{equation}
Finally, using again the expression in \cref{eq:wave_vector} for
$\kappa(\omega,\fluxC)$ we obtain the expression for the phase shift $\phi$ as a
function of the system parameters. 
\begin{equation}
\tan{\phi(\omega,\fluxT,\fluxC)} =
\left[1-\frac{2\Ceff(\fluxT,\omega)}{\Cg}\right]\sqrt{\frac{\Cg}{\Cg+4\Cj}}\frac{1}{\sqrt{\left(\frac{\wplasma(\fluxC)}{\omega}\right)^2-1}}.
\label{eq:phi_analytic_2}
\end{equation}

\subsection{Link between the scattering phase shift and the relative frequency shift}

In the main text we defined the relative frequency shift $\delta\phi_n$ in terms
of the discrete mode frequencies of the full  system (transmon plus finite chain
of $N$ nodes) at respective transmon fluxes $\fluxT$ and $\flux_0/2$. We repeat
the definition here:
\begin{equation}
\delta\phi_n(\fluxT,\fluxC)=\pi\frac{\omega_\mode(\flux_0/2,\fluxC)-\omega_\mode(\fluxT,\fluxC)}{\omega_\mode(\flux_0/2,\fluxC)-\omega_{\mode-1}(\flux_0/2,\fluxC)}.
\label{eq:exp_TPS}
\end{equation} 
Here we relate this to the relative scattering phase shift
$\delta\phi(\omega,\fluxT,\fluxC)$ for the infinite system (see Eq.\,14 in the {\it Methods} section of the 
main text)
by showing that
\begin{equation}
\delta\phi_n(\fluxT,\fluxC)=\delta\phi(\omega_\mode(\fluxT,\fluxC),\fluxT,\fluxC)
+\mathcal O(N^{-1}).
\label{eq:rel_freq_shift_scat_shift}
\end{equation}
For conveniece we assume that  island $\Nj+1$ is grounded. (The precise boundary condition becomes
immaterial in the $N\to\infty$ limit.) Had the chain been open to the left of
node $1$, the eigenmodes would have been
$\varphi_\island\propto\cos\left(\kappa^{0}_\mode\lencell\island\right)$,
$\island \in 1,2,3,\ldots,\Nj$ with wave numbers given by $\kappa^{0}_\mode =
(\mode-1/2)\pi/\Nj\lencell$ with $\mode \in 1,2,3,\ldots,\Nj$. In the presence
of the transmon to the left of chain node $1$, the eigenmodes inside the chain
change to $\cos\left(\kappa_\mode\lencell\island - \phi_\mode\right)$. Here
$\phi_\mode$ is the additional phase introduced by the transmon. Now the
$\kappa_\mode$ depend on $\phi_\mode$ too. Assuming the boundary conditions that
the nodes to the left of transmon island $L$ and to the right of chain island
$\Nj$ are grounded, they are given by
\begin{equation}
\kappa_\mode\lencell{} = \frac{\left(\mode-\frac{1}{2}\right)\pi}{N}
+\frac{\phi_\mode}{N} = \kappa^{0}_\mode\lencell{} +\frac{\phi_\mode}{N}.
\end{equation} 
The modes of the system follow a dispersion relation
$\omega_\mode\left(\fluxT,\fluxC\right) = \omega(\fluxC,\kappa_\mode)$. The
notation must be understood as follows: $\omega_\mode(x,y)$ and $\omega(x,y)$
denote distinct functions. The two arguments of the former refer to respectively
the flux in a transmon and in a chain SQUID, and for given fluxes, the function
assumes the value of the frequency of system mode $n$. The first argument of the
latter function $\omega(x,y)$ refers to the flux in a chain SQUID, while the
second argument refers to the unquantized wave number of a mode in the infinite
chain. The function evaluates to the frequency corresponding to the given wave
number (which does not depend on the transmon flux). For sufficiently large
$\Nj$ we can expand the dispersion relation around $\kappa^0_\mode\lencell{}$
\begin{equation}
\omega_\mode\left(\fluxT,\fluxC\right) =
\omega\left(\fluxC,\kappa^0_\mode\right)+\frac{\phi_\mode\left(\fluxT,\fluxC\right)}{\Nj}\left.\frac{\partial\omega(\fluxC,\kappa)}{\partial
\left(\kappa\lencell{}\right)}\right|_{\kappa=\kappa_\mode^{0}},
\end{equation}
with corrections of order $\Nj^{-2}$. The dependence on the transmon and chain fluxes is included. Similarly, we can expand $\omega_{\mode-1}\left(\fluxT,\fluxC\right)$ around $\kappa^{0}_\mode\lencell{}$ to obtain
\begin{equation}
\omega_{\mode-1}\left(\fluxT,\fluxC\right)=\omega_\mode\left(\fluxT,\fluxC\right)-\frac{\pi}{\Nj}\left.\frac{\partial\omega(\fluxC,\kappa)}{\partial\left( \kappa \lencell{}\right)}\right|_{\kappa=\kappa_\mode^{0}}.
\end{equation}
Here we made use of the fact that $\phi_\mode - \phi_{\mode-1} = \mathcal{O}\left(\Nj^{-1}\right)$. Therefore we can set $\phi_{\mode-1} = \phi_\mode$ introducing an error of $\mathcal{O}\left(\Nj^{-2}\right)$ which can be ignored for large $\Nj$. We can now obtain the terms in \cref{eq:exp_TPS},  
\begin{align}
&\omega_\mode\left(\flux_0/2,\fluxC\right) = \omega\left(\fluxC,\kappa^0_\mode\right)+\frac{\phi_\mode\left(\flux_0/2,\fluxC\right)}{\Nj}\left.\frac{\partial\omega(\fluxC,\kappa)}{\partial \left(\kappa\lencell{}\right)}\right|_{\kappa=\kappa_\mode^{0}}\\
&\omega_\mode\left(\fluxT,\fluxC\right) = \omega\left(\fluxC,\kappa^0_\mode\right)+\frac{\phi_\mode\left(\fluxT,\fluxC\right)}{\Nj}\left.\frac{\partial\omega(\fluxC,\kappa)}{\partial \left(\kappa\lencell{}\right)}\right|_{\kappa=\kappa_\mode^{0}}\\
&\omega_{\mode-1}\left(\flux_0/2,\fluxC\right) =
\omega_\mode\left(\flux_0/2,\fluxC\right)-\frac{\pi}{\Nj}\left.\frac{\partial\omega(\fluxC,\kappa)}{\partial
\left(\kappa\lencell{}\right)}\right|_{\kappa=\kappa_\mode^{0}}.
\end{align}
Substituting this into \cref{eq:exp_TPS}, and noting that 
$\phi_\mode\left(\fluxT,\fluxC\right)=\phi\left(\omega_\mode(\fluxT,\fluxC),\fluxT,\fluxC\right) $, we obtain \cref{eq:rel_freq_shift_scat_shift}.

\subsection{Relation between the Phase Shift and the impurity response function}

\begin{figure}[ht]
\begin{center}
\includegraphics[scale=0.4]{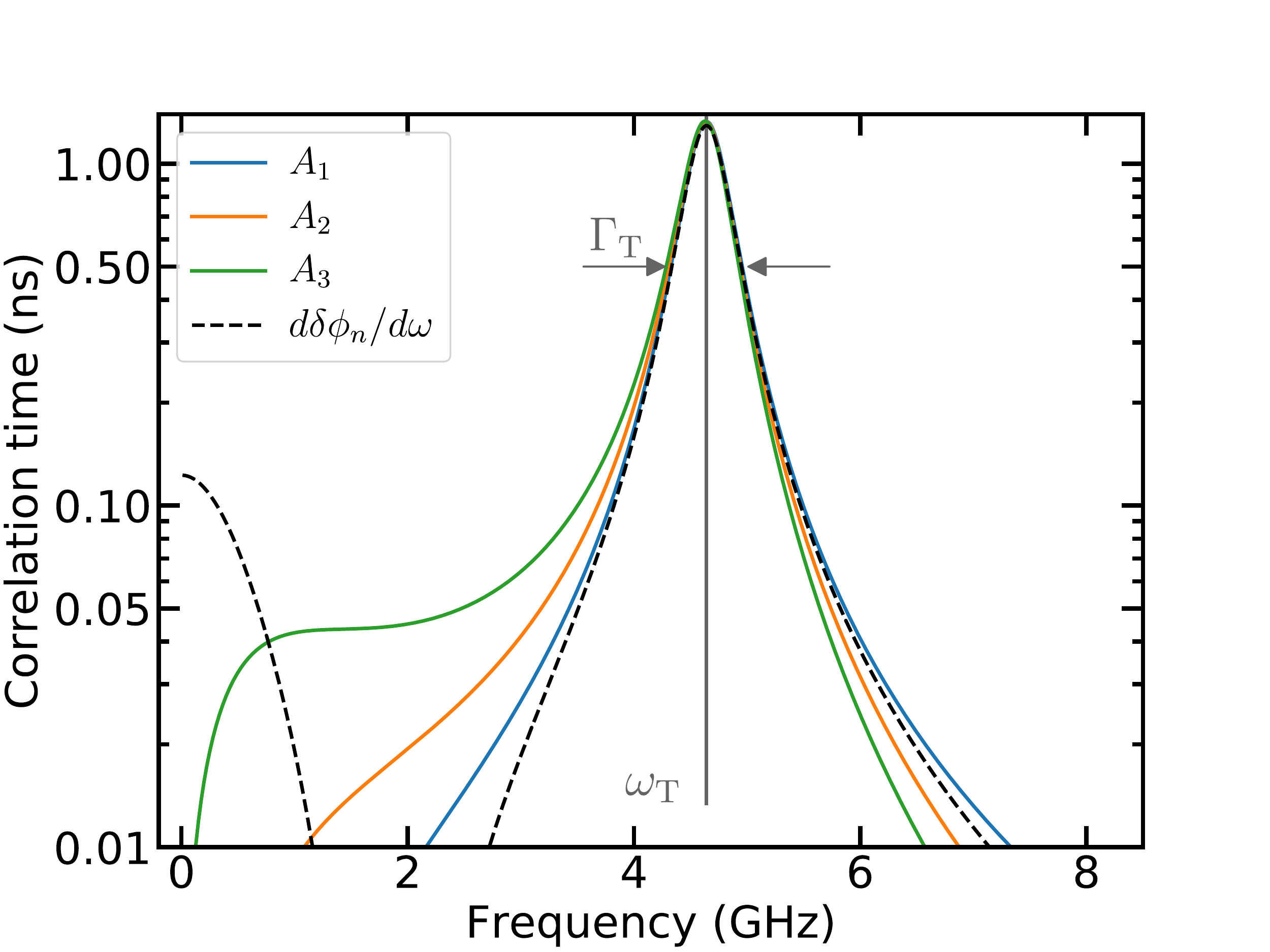}
\caption{{\bf Connection between phase shifts and qubit dissipation.} Comparison between the 
three correlation functions and the energy derivative of the phase shift.}
\label{fig:correlation_TPS}
\end{center}
\end{figure}

To elaborate the link between the local impurity response function and the phase
shift induced by the transmon qubit, we define three spectral densities
\begin{align}
A_1(\omega) &=
\frac{2\Eself(\fluxT)}{\Qfreq{}}\text{Re}\int_0^{\infty}\frac{dt}{2\pi}e^{i\omega
t}\left\langle [\phaseop_\text{T}(t),\phaseop_\text{T}(0)]\right\rangle, \\
A_2(\omega) &= -\text{Im}\int_0^{\infty}\frac{dt}{2\pi}e^{i\omega t}\left\langle
[\phaseop_\text{T}(t),\nop_\text{T}(0)]\right\rangle, \\
A_3(\omega) &= \frac{2\Qfreq{}}{\Eself(\fluxT)}\text{Re}\int_0^{\infty}\frac{dt}{2\pi}e^{i\omega t}\left\langle [\nop_\text{T}(t),\nop_\text{T}(0)]\right\rangle,
\end{align}
corresponding to the phase-phase, phase-charge and charge-charge response of the
transmon up to constant prefactors. In order to calculate these spectral
densities we turn to the quantum mechanical problem (In the previous section, we
could compute the phase shift by solving the classical equations of motion).
With each mode $\omega$ we associate cannonical bosonic operators $b_{\omega}$
and $b_{\omega}^{\dagger}$. These are related to the charge operator
$\nop_{\island}$ and phase operator $\phaseop_{\island}$ for each of the islands
as
\begin{align}
\phaseop_{\island} &=
\frac{-i}{\sqrt{2}}\int_{0}^{\wplasma(\fluxC)}d\omega\varphi_{\island}(\omega)\left(b_{\omega}
- b_{\omega}^{\dagger}\right), \\
\nop_{\island} &=
\frac{1}{\sqrt{2}}\int_{0}^{\wplasma(\fluxC)}\frac{d\omega}{\omega}\sum_{\iisland=L,R,1,2,\cdots}
\Jm(\fluxT,\fluxC)_{\island,\iisland}\varphi_{\iisland}(\omega)\left(b_{\omega}
+ b_{\omega}^{\dagger}\right),
\end{align}
where the profile $\varphi_{\island}(\omega)$ is normalized such that
\begin{equation}
\sum_{\island = L,R,1,2,\cdots}
\varphi_{\island}(\omega)\Jm(\fluxT,\fluxC)_{\island,\iisland}\varphi_{\iisland}(\omega')
= \omega\delta(\omega - \omega').
\end{equation}
The normalization constant in \cref{eq:cosine_mode} is thus set to
\begin{equation}
N_{\omega} = \sqrt{\frac{\hbar\omega}{\pi\Ej(\fluxC)[1-\cos
\kappa(\omega,\fluxC)]}\frac{\partial\kappa(\omega,\fluxC)}{\partial\omega}}.
\end{equation}
From the definition of $\nop_\text{T}$ and $\phaseop_\text{T}$ (see the text below Eq.\,7 in the {\it Methods} section of the main text) follows
\begin{eqnarray}
\nop_\text{T} &=& (\nop_R-\nop_L)/2 
 =
\frac{1}{\sqrt{2}}\int_{0}^{\wplasma(\fluxC)}d\omega\frac{\Eself(\fluxC)}{\omega}\left[\varphi_R(\omega)
- \varphi_L(\omega)\right]\left(b_{\omega} + b_{\omega}^{\dagger}\right),\\
\phaseop_\text{T} & = &\phaseop_R-\phaseop_L
 = \frac{-i}{\sqrt{2}}\int_{0}^{\wplasma(\fluxC)}d\omega\left[\varphi_R(\omega) -
\varphi_L(\omega)\right]\left(b_{\omega} - b_{\omega}^{\dagger}\right).
\end{eqnarray}
Explicitly, we find:
\begin{equation}
\varphi_R(\omega) - \varphi_L(\omega) = N_{\omega}\frac{\Cc\omega^2/(2e)^2}{(C_0
+ \CshQ)\omega^2/(2e)^2-2\Eself(\fluxT)}\cos \phi(\omega,\fluxT,\fluxC).
\end{equation}
In the Heisenberg picture $b_{\omega}(t) = e^{i\omega t}b_{\omega}$ and
$b_{\omega}^\dagger(t) = e^{-i\omega t}b_{\omega}^{\dagger}$. Using this to
calculate the spectral densities $A_j(\omega)$ for $\omega > 0$ we obtain
\begin{align}
A_1(\omega) &= \frac{\Eself(\fluxT)}{\Qfreq{}}\left[\varphi_R(\omega) -
\varphi_L(\omega)\right]^2,\\
A_2(\omega) &= \frac{\Eself(\fluxT)}{\omega}\left[\varphi_R(\omega) -
\varphi_L(\omega)\right]^2,\\
A_3(\omega) &= \frac{\Eself(\fluxT)\Qfreq{}}{\omega^2}\left[\varphi_R(\omega) - \varphi_L(\omega)\right]^2
\end{align} 
Now we compare the three correlation functions with the frequency derivative of
$\delta\phi(\omega,\fluxT,\fluxC)=\phi(\omega,\fluxT,\fluxC)-\phi(\omega,\flux_0/2,\fluxC)$ with 
$\phi(\omega,\fluxT,\fluxC)$
given in \cref{Seq:phi_analytic}. In \cref{fig:correlation_TPS} we plot the four
curves. We see that the four curves overlap around the transmon frequency
$\Qfreq{}$. This means that the width and the center frequency obtained from the
scattering phase shift are good estimations of the real width $\Qwidth{}$ and
frequency $\Qfreq{}$ of the transmon.

\subsection{Breakdown of the rotating wave approximation}

In this section, we investigate the applicability of the rotating wave approximation (RWA), a common technique for analysing the light-matter interaction at sufficiently weak coupling. 
The regime in which the light-matter coupling is so large that this approximation becomes inaccurate, is referred to as ultra-strong coupling.
We will find that indeed, the RWA leads to errors of a few percent for our device. 

To set up the RWA, we have to identify the harmonic oscillator basis that diagonalizes the (finite) chain part of
the Hamiltonain. For this purpose, it is convenient to define $N\times N$ matrices ${\widehat C^{-1}}_{\rm chain}$ and ${\widehat L^{-1}}_{\rm chain}$ with entries
\begin{equation}
[\widehat C^{-1}_{\rm chain}]_{jk}=[\widehat C^{-1}]_{jk},~~~[\widehat L^{-1}_{\rm chain}]_{jk}=[\widehat J]_{jk}/\varphi_0^2,~~~j,\,k=1,\,2,\,\ldots,\,N,
\end{equation}
i.e. ${\widehat C^{-1}}_{\rm chain}$ and $\varphi_0^2{\widehat L^{-1}}_{\rm chain}$ are the lower right $N\times N$ blocks of respectively the inverse of the full $(N+2)\times(N+2)$ capacitance matrix $\widehat C$, and of the full $(N+2)\times(N+2)$ Josephson matrix $\widehat J$. We then define an $N\times N$ matrix $\widehat \Pi_{\rm chain}$ and a positive definite
diagonal matrix $\widehat \omega_{\rm chain}$ such that the columns of $\widehat \Pi_{\rm chain}$ contain the eigenvectors of $\widehat C^{-1}_{\rm chain} \widehat L^{-1}_{\rm chain}$ while the
diagonal entries of $(\widehat \omega_{\rm chain})^2$ are the corresponding eigenvalues, i.e.
\begin{equation}
\widehat C^{-1}_{\rm chain} \widehat L^{-1}_{\rm chain}\widehat \Pi_{\rm chain}=\widehat \Pi_{\rm chain}(\widehat \omega_{\rm chain})^2.
\end{equation}
Since the eigenvalues are real, we can and will choose the entries of $\widehat \Pi_{\rm chain}$ to be real as well. 
This definition determines each column of $\widehat \Pi_{\rm chain}$ up to a normalization constant. We fix these constants by demanding that
\begin{equation}
\varphi_0^2\sum_{jk} [\widehat L^{-1}_{\rm chain}]_{jk}[\widehat \Pi_{\rm chain}]_{jl}[\widehat \Pi_{\rm chain}]_{kl}=\hbar [\omega_{\rm chain}]_l,\label{norm}
\end{equation}
where $[\omega_{\rm chain}]_l$ is the $l$'th diagonal entry of $\widehat \omega_{\rm chain}$. We also define a matrix 
\begin{equation}
\widehat \Xi_{\rm chain}=\varphi_0^2 \widehat L^{-1}_{\rm chain} \widehat \Pi_{\rm chain} \widehat \omega_{\rm chain}^{-1}/\hbar.
\end{equation}
It is easy to verify that the row $l$ of $\widehat \Xi_{\rm chain}^{\rm T}$ contains the left-eigenvector of $\widehat C^{-1}_{\rm chain} \widehat L^{-1}_{\rm chain}$ that
is associated with eigenvector $ ([\omega_{\rm chain}]_l)^2$. As a result $\widehat \Xi_{\rm chain}^{\rm T} \widehat \Pi_{\rm chain}$ is guaranteed to be a diagonal matrix. 
Furthermore, due to the normalization condition (\ref{norm}) we chose for $\widehat \Pi_{\rm chain}$, the diagonal entries of $\widehat \Xi_{\rm chain}^{\rm T} \widehat \Pi_{\rm chain}$ are all equal to unity. Thus
\begin{equation}
\widehat \Xi_{\rm chain}^{\rm T}=\widehat \Pi_{\rm chain}^{-1}.\label{inverse}
\end{equation}
We now define $N$ operators
\begin{equation}
\widehat b_{{\rm chain},k}=\frac{1}{\sqrt{2}}\sum_{j=1}^N\left\{\widehat n_j [\widehat \Pi_{\rm chain}]_{jk}+i \widehat \varphi_j [\widehat \Xi_{\rm chain}]_{jk}\right\}.
\end{equation}
Owing to (\ref{inverse}) and the fact that $\left[\widehat n_j,\widehat \varphi_k\right]=i\delta_{j,k}$, the operators $\widehat b_{{\rm chain},k}$, $k=1,\,2,\,\ldots,\,N$ are bosonic annihilation operators,
i.e. $[\widehat b_{{\rm chain},j},\widehat b_{{\rm chain},k}]=0$ and $[\widehat b_{{\rm chain},j},\widehat b_{{\rm chain},k}^\dagger]=\delta_{j,k}$. Furthermore, for the chain part
of the Hamiltonian we obtain
\begin{align}
H_{\rm chain}&=\frac{1}{2}\sum_{j,k=1}^N \left\{(2e)^2 [\widehat C^{-1}_{\rm chain}]_{jk}\widehat n_j \widehat n_k+\varphi_0^2 [\widehat L^{-1}_{\rm chain}]_{jk}\widehat \varphi_k
\widehat \varphi_j\right\}\nonumber\\
&=\hbar \sum_{j=1}^N [\omega_{\rm chain}]_j \left(\widehat b_{{\rm chain},j}^\dagger \widehat b_{{\rm chain},j}-\frac{1}{2}\right).
\end{align}
For the term in the Hamiltonian that couples the transmon to the chain, we find
\begin{align}
H_{\rm coupling}&=\widehat n_{\rm T}\sum_{j=1}^N \nu_j \widehat n_j\nonumber\\
&=\frac{1}{\sqrt{2}} \widehat n_{\rm T} \sum_{k=1}^N g_k \left(\widehat b_{{\rm chain},k}+ \widehat b_{{\rm chain},k}^\dagger\right) 
\end{align} 
where
\begin{equation}
g_k=\sum_{j=1}^N \nu_j [\widehat \Xi_{\rm chain}]_{jk}.
\end{equation}

A standard way to proceed from here is to truncate the full Hilbert space of the transmon to the subspace spanned by two lowest energy eigenstates $\left|0_{\rm T}\right>$
and $\left|1_{\rm T}\right>$ of the isolated transmon Hamiltonian (\ref{eq:isolated_transmon_hamiltonain}). This leads to a Hamiltonian of the Jaynes-Cummings type, ubiquitous in  Quantum Optics. 
At sufficiently weak coupling, the expectation is that this should be accurate for studying the situation where a near-resonant excitation from the chain induces a transition between
the ground and first excited states of the transmon. The operator 
$\widehat n_{\rm T}$ is replaced by
\begin{equation}
\widehat n_{\rm T}\simeq \left<0_{\rm T}\right|\widehat n_{\rm T}\left|1_{\rm T}\right>\left\{\left|0_{\rm T}\right>\left<1_{\rm T}\right|+\left|1_{\rm T}\right>\left<0_{\rm T}\right|\right\},
\end{equation}
where we've chosen the overall phases of $\left|0_{\rm T}\right>$
and $\left|1_{\rm T}\right>$ such that $\left<0_{\rm T}\right|\widehat n_{\rm T}\left|1_{\rm T}\right>$ is real.

Alternatively, a more controlled way to proceed is to make the self-consistent harmonic approximation (SCHA) (see Sec. \ref{sec:scha}), which we have shown to be well-justified, and to express $\widehat n_{\rm T}$ in terms of the resulting bosonic 
transmon operators [see Eq. (\ref{eq:Bdef})], i.e.
\begin{equation}
\widehat n_{\rm T}=\frac{1}{\sqrt{2}}\left(\frac{E_{\rm C,T}}{E_{\rm S}}\right)^{1/4}(B+B^\dagger).
\end{equation}
After the SCHA, the Hamiltonian becomes quadratic, and no further approximations are required. We will however still consider the effect of making the RWA on this quadratic Hamiltonian, in order to asses whether or not the assumptions underpinning the RWA are valid in our device.

The RWA approximation now involves dropping the transmon-chain coupling terms in which the transmon (in the unperturbed basis) is excited while a boson is emitted into the chain,
or the transmon is de-exited while a boson is absorbed from the chain. We adopt the standard nomenclature and refer to the dropped terms as ``counter-rotating'' (based on their time-dependence in the Dirac picture). Depending on whether this approximation is made in conjunction with truncating the transmon Hilbert space or with the SCHA, we either obtain
an RWA Hamiltonian
\begin{align}
H_{{\rm RWA},1}&=(E_{1,{\rm T}}-E_{0,{\rm T}})\left|1_{\rm T}\right>\left<1_{\rm T}\right|+\sum_{n=1}^N \hbar [\omega_{\rm chain}]_n \widehat b_{{\rm chain},n}^\dagger 
\widehat b_{{\rm chain},n}\nonumber\\
&+\frac{1}{\sqrt{2}} 
 \left<0_{\rm T}\right|\widehat n_{\rm T}\left|1_{\rm T}\right>
 \sum_{k=1}^N g_k  
 \left(\left|0_{\rm T}\right>\left<1_{\rm T}\right|\widehat b_{{\rm chain},k}+\left|1_{\rm T}\right>\left<0_{\rm T}\right| \widehat b_{{\rm chain},k}^\dagger\right),
\end{align}
or 
\begin{align}
H_{{\rm RWA},2}&=\omega_{\rm T}B^\dagger B+\sum_{n=1}^N \hbar [\omega_{\rm chain}]_n \widehat b_{{\rm chain},n}^\dagger 
\widehat b_{{\rm chain},n}+\frac{1}{2} \left(\frac{E_{\rm C,T}}{E_{\rm S}}\right)^{1/4}
 \sum_{k=1}^N g_k  \left(B^\dagger \widehat b_{{\rm chain},k}+\widehat b_{{\rm chain},k}^\dagger B\right).
\end{align}
For both Hamiltonians, the ground state is trivial: the transmon is in its unperturbed ground state, and there are no bosonic excitations in the chain. We measure
energy relative to this ground state. Both Hamiltonians leave invariant
the subspace spanned by states in which there are no bosons in the chain, while the transmon is in its unperturbed first excited state, or there is one boson in the chain while the 
transmon is in its unperturbed ground state. The excited states relevant for spectroscopy at low driving power are found by diagonalizing the RWA Hamiltonians in this subspace.

\begin{figure}[ht]
\begin{center}
\includegraphics[width=.7\textwidth]{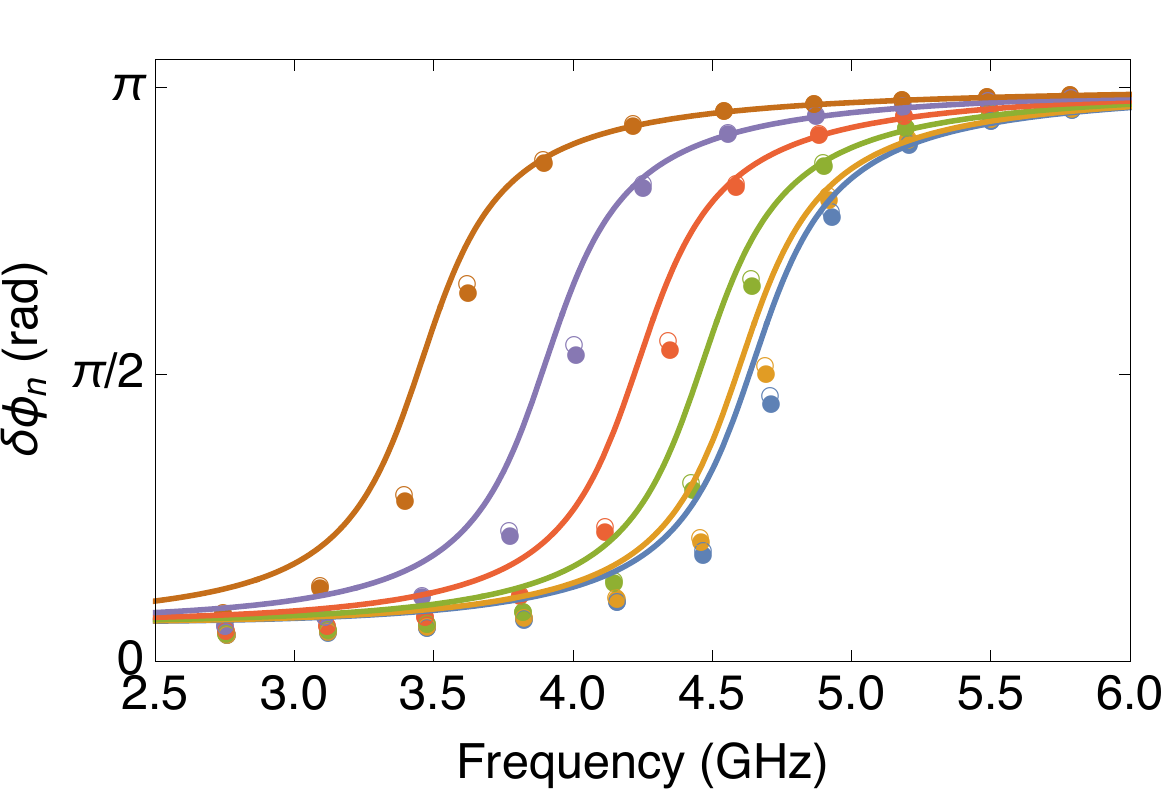}
\caption{{\bf Deviations from the microscopic model under the RWA assumption.} Curves show the analytical formula for the relative frequency shift (Eq. 14 in the main text). Open circles show the corresponding result calculated using $H_{{\rm RWA},1}$
and closed circles the result calculated using $H_{{\rm RWA},2}$. In all cases the system parameters were taken as in Table 1 of the main text. The flux $\Phi_{\rm C}$ through chain 
SQUIDS was held fixed at zero. Results for six different fluxes $\Phi_T\in[0,0.3\Phi_0]$ through the transmon SQUID are shown.}
\label{fig:rwa}
\end{center}
\end{figure}

In Figure \ref{fig:rwa} we compare the relative frequency shift (Eq. 1 in the main text) predicted by $H_{\rm RWA, 1}$ and $H_{\rm RWA, 2}$ to the analytical SCHA formula (Eq. 14 in the 
main text) derived for an infinite chain. We have also computed SCHA results for the finite chain of 4700 islands, and found that they lie on top of the infinite chain curves. We omit them from the figure to avoid clutter. We note that $H_{{\rm RWA},1}$ and $H_{{\rm RWA},2}$ give very similar results. This is consistent with our claim that at low energies, the SCHA Hamiltonian from which $H_{{\rm RWA},2}$ derives, is a good approximation to the full Hamiltonian from which $H_{{\rm RWA},1}$ is derived. We ascribe the small difference between
results for $H_{{\rm RWA},1}$ and $H_{{\rm RWA},2}$ to the truncation by hand in $H_{{\rm RWA},1}$ of the transmon Hilbert space to two states. (No such by-hand truncation was required in $H_{{\rm RWA},2}$.) If we fit an arctan line shape (Eq. 18 in the main text) to the SCHA curves in the figure, we find that the transmon resonance occurs at a frequency within about $0.01$ GHz from $h^{-1}$ times the energy difference between the ground and first excited states of the isolated transmon. Using the same procedure on the relative frequency shift predicted by either $H_{{\rm RWA},1}$ or $H_{{\rm RWA},2}$ on the other hand, gives a resonance frequency that is $\sim 0.1$ GHz higher than $h^{-1}$ times the ground to first excitation energy of the isolated transmon. We conclude that the RWA approximation is not quantitatively accurate, producing an error of between 2\% and 5\% for the transmon resonance frequency. This signals that our device indeed operates in the ultra-strong light-matter coupling regime.

\end{document}